\documentclass[apj, letterpaper]{emulateapj}  

\pdfoutput=1

\usepackage{amsmath}

\def\z{\tilde z}

\begin{document}

\slugcomment{\bf}
\slugcomment{ApJ, in press}

\title{Atmospheric dynamics of brown dwarfs and directly imaged 
giant planets}

\shorttitle{Atmospheric dynamics of brown dwarfs}
\shortauthors{Showman and Kaspi}

\author{Adam P. Showman\altaffilmark{1}, Yohai Kaspi\altaffilmark{2}}

\altaffiltext{1}{Department of Planetary Sciences and Lunar and Planetary
Laboratory, The University of Arizona, 1629 University Blvd., Tucson, AZ 85721 USA; 
showman@lpl.arizona.edu}
\altaffiltext{2}{Center for Planetary Science, Weizmann Institute of Science, Rehovot, Israel}

\begin{abstract}
\label{Abstract}
A variety of observations now provide evidence for vigorous motion in
the atmospheres of brown dwarfs and directly imaged giant planets.
Motivated by these observations, we examine the dynamical regime of
the circulation in the atmospheres and interiors of these objects.
Brown dwarfs rotate rapidly, and for plausible wind speeds, the flow
at large scales will be rotationally dominated, exhibiting geostrophic
balance between pressure-gradient and Coriolis forces.  We present
three-dimensional, global, anelastic numerical simulations of
convection in the interior.  Fundamental theory, scaling arguments,
and our anelastic simulations all demonstrate that, at large scales,
the convection aligns in the direction parallel to the rotation axis.
Convection occurs more efficiently at high latitudes than low
latitudes, leading to systematic equator-to-pole temperature
differences that may reach $\sim$1 K near the top of the convection
zone. The rotation significantly modifies the convective properties.
The interaction of convection with the overlying, stably stratified
atmosphere will generate a wealth of atmospheric waves, and we argue
that, just as in the stratospheres of planets in the solar system, the
interaction of these waves with the mean flow will lead to a
significant atmospheric circulation at regional to global scales.  At
scales exceeding thousands of km, this should consist of
geostrophically balanced, stratified turbulence (possibly
organizing into coherent structures such as vortices and jets) and an
accompanying overturning circulation.  We present a semi-quantitative,
analytic theory of this circulation as a function of the wave-driving
efficiency.  For plausible efficiencies, this theory predicts
characteristic horizontal temperature variations of several to
$\sim$50 K, horizontal wind speeds of $\sim$10--$300\rm\,m\,s^{-1}$,
and vertical velocities that advect air over a scale height in
$\sim$$10^5$--$10^6\rm\,s$.  The vertical mixing associated with this
large-scale circulation may help to explain the chemical quenching of
CO and NH$_3$ observed on some brown dwarfs.  Moreover, the implied
large-scale organization of temperature perturbations and vertical
velocities suggests that, near the L/T transition, patchy clouds
can form near the photosphere, helping to explain recent observations
of brown-dwarf variability in the near-infrared.

\end{abstract}

\keywords{planets and satellites: general, planets and satellites: 
individual: HD 209458b, methods: numerical, atmospheric effects}


\section{Introduction}
\label{Introduction}

Since the discovery of brown dwarfs beginning in the mid-1990s, our
understanding of the atmospheric structure of these objects has grown
ever more sophisticated.  Approximately 1000 brown dwarfs, and a handful of
directly imaged planets, have now been
discovered.  Observational acquisition of infrared (IR) spectra for
many of these objects have allowed the definition of the L, T, and Y
spectral classes \citep[e.g.,][]{kirkpatrick-2005, cushing-etal-2011}.
The theory for these objects now encompasses a broad understanding
of their evolution, radii, luminosity,
molecular composition, spectra, and colors, and includes prescriptions
for condensate formation and rainout, surface patchiness, and
disequilibrium chemistry.     Notably, however, these theoretical
advances have relied heavily on one-dimensional (1D) models 
for the atmospheric radiative transfer and interior evolution
\citep[for an early review, see][]{burrows-etal-2001}.  By
comparison, little effort has been made to understand the 
global, three-dimensional atmospheric dynamics of these substellar bodies.

Yet there is increasing evidence that brown dwarfs exhibit vigorous
atmospheric circulations.  This evidence falls into three main
classes.  First, L dwarfs, particularly of later spectral type, show a
reddening of near-infrared (e.g., $J-K$) colors that indicate the
presence of silicate clouds in the visible atmospheres
\citep[e.g.,][]{kirkpatrick-etal-1999, kirkpatrick-2005,
  chabrier-etal-2000, tsuji-2002, cushing-etal-2006, knapp-etal-2004}.
Since cloud particles would gravitationally settle in the absence of
dynamics, such clouds imply the presence of atmospheric vertical
mixing necessary to keep the particles suspended.  In the cooler T
dwarfs, the condensation occurs progressively deeper and, for objects
with sufficiently low effective temperature, eventually no longer
influences the infrared spectrum.  However, the L/T transition itself
remains poorly understood; it occurs over a surprisingly small range
of effective temperature and accompanies a $J$-band brightening that
are not easily captured by standard 1D models
\citep{chabrier-etal-2000, allard-etal-2001, burrows-etal-2006,
  saumon-marley-2008}.  Hypotheses that have been put forward to
resolve this discrepancy are that, across the transition, the cloud
sedimentation efficiency changes \citep{knapp-etal-2004} or that the
clouds become patchy, allowing contributions from both cloudy and
cloud-free regions to affect the disk-integrated emergent spectrum
\citep{burgasser-etal-2002, marley-etal-2010}.  In both cases, a role
for atmospheric dynamics in modulating the clouds is implicated.

A second line of evidence for atmospheric circulation comes from 
chemical disequilibrium of CO, CH$_4$, and NH$_3$ inferred for
many cool brown dwarfs. Late T dwarfs have
sufficiently cool atmospheres that the preferred chemical-equilibrium
forms of carbon and nitrogen near the photosphere are CH$_4$ and NH$_3$, 
respectively; in contrast, CO and N$_2$ dominate 
under the high-pressure and temperature regions 
at depth.  Fitting of infrared (IR) spectra to 
radiative transfer models shows that, near the photosphere, many T dwarfs
exhibit an overabundance of CO and an underabundance of NH$_3$
relative to chemical equilibrium.  This can be attributed to vertical
transport of CO-rich and NH$_3$-poor air from depth and the subsequent
chemical quenching of these disequilibrium mixing ratios due to the
long chemical interconversion timescales in the low-pressure, low-temperature
regions near the photosphere.  This story was first worked out
for CO on Jupiter \citep{prinn-barshay-1977, bezard-etal-2002,
visscher-moses-2011} and then for both CO and NH$_3$ on
Gl 229b \citep{fegley-lodders-1996, noll-etal-1997, griffith-yelle-1999,
saumon-etal-2000}.  Subsequently, chemical disequilibrium and vertical
mixing have been inferred in the atmospheres of a wide range of T 
dwarfs \citep{saumon-etal-2006,
saumon-etal-2007, hubeny-burrows-2007, leggett-etal-2007, leggett-etal-2007b, 
leggett-etal-2008, leggett-etal-2010, stephens-etal-2009}.

Third, recent near-IR photometric observations demonstrate that
several brown dwarfs near the L/T transition exhibit large-amplitude
variability over rotational timescales, probably due to cloudy and
relatively cloud-free patches rotating in and out of view.  The
possibility of weather on brown dwarfs has long motivated searches for
variability.  Recently, \citet{artigau-etal-2009} observed the T2.5
dwarf SIMP0136 in $J$ and $K_s$ band and found peak-to-peak
modulations of $\sim$5\% ($\sim$50 mmag) throughout the inferred
2.4-hr rotation period.  \citet{radigan-etal-2012} observed the T1.5
dwarf 2M2139 in $J$, $H$, and $K_s$ and found peak-to-peak variations
of up to $\sim$25\% with an inferred rotation period of either
7.7 or 15.4 hr.  The relative amplitudes of the variability at $J$,
$H$, and $K_s$ place strong constraints on the cloud and thermal
structure associated with the variability.  These authors considered
models where the variability resulted from lateral variations in
effective temperature alone (with no variations in the cloud
properties), lateral variations in the cloud properties alone (with no
variation in effective temperature), and lateral variations in both
temperature and cloud properties.  The observations rule out models
with a uniform cloud deck and instead strongly favor models with
significant lateral variations in both cloud opacity and effective
temperature; the relatively cloud-free regions exhibit effective
temperatures $\sim$100--400 K greater than the cloudier regions.  This
suggests a picture with spatially distinct regions of lower and higher
condensate opacity, where radiation escapes to space from
lower-pressure, cooler levels in the high-opacity regions and deeper,
warmer levels in the low-opacity regions.  The observations of
\citet{artigau-etal-2009} and \citet{radigan-etal-2012} both show that
the lightcurves vary significantly over intervals of several Earth
days, indicating that the shape, orientation, or relative positions of
the low- and high-condensate opacity regions evolve over timescales of
days.

In addition to these observations of field brown dwarfs,
growing numbers of young, hot extrasolar giant planets (EGPs) 
are being imaged and characterized.   Prominent discoveries include
planetary-mass companions to $\beta$ Pic, 2M1207, and HR 8799.  
Multi-band photometry already
indicates that 2M1207b and several of the HR 8799 planets
exhibit clouds and probably disequilibrium chemistry similar to
that inferred on brown dwarfs \citep{hinz-etal-2010,
bowler-etal-2010, currie-etal-2011,
galicher-etal-2011, skemer-etal-2011, madhusudhan-etal-2011b,
barman-etal-2011, barman-etal-2011b, marley-etal-2012}.  
With effective temperatures
exceeding $\sim$$1000\rm\,K$, these planets radiate 
IR fluxes $\gtrsim 10^5\rm\,W\,m^{-2}$, orders of magnitude
greater than the flux received by their primary star.  Stellar 
irradiation is therefore negligible to their dynamics.  From a 
meteorology perspective, this population of bodies will therefore 
resemble low-mass, low-gravity versions of free-floating brown dwarfs.
With next-generation telescope facilities, including the Gemini
Planet Imager and SPHERE, significant numbers of new planets will 
be discovered, greatly opening our ability to probe planetary
meteorology at the outer edge of stellar systems.

These existing and upcoming observations provide strong motivation for
investigating the global atmospheric dynamics of brown dwarfs and
directly imaged planets.  As yet, however, no investigations of the
global atmospheric circulation of brown dwarfs have been performed.
The only study of brown dwarf atmospheric dynamics published to date is that of
\citet{freytag-etal-2010}, who performed two-dimensional, non-rotating
convection simulations in a box $\sim$$400\rm\,km$ wide by
$\sim$$150\rm\,km$ tall.  Their study provides valuable insights into
the role of convectively generated small-scale gravity waves in
causing vertical mixing.  Nevertheless, dynamics on scales of tens to
hundreds of km differs substantially from that on global scales of
$10^4$--$10^5\,$km, and thus, for understanding the global-scale
circulation---including the implications for variability---it is
essential to consider global-scale models.

Here, we aim to fill this gap by presenting the first global-scale
models of brown dwarf atmospheric dynamics.  Rotation periods of L and
T dwarfs inferred from spectral line broadening range from $\sim$2 to
12 hours \citep{zapatero-osorio-etal-2006, reiners-basri-2008}, in
line with the periods of SIMP0136 and 2M2139 inferred from lightcurve
modulation \citep{artigau-etal-2009, radigan-etal-2012}.  We will show
that, at these periods, rotation dominates the global-scale dynamics
and will constitute an overriding factor in controlling the
three-dimensional wind and temperature structure.  We first present
theoretical arguments to highlight the fundamental dynamical regime in
which brown dwarfs lie, to show how rotation organizes the large-scale
wind, establishes systematic temperature differences, and shapes the
convective velocities (Section~\ref{theory}).  We next present global,
three-dimensional numerical simulations of the interior convection of
brown dwarfs that confirm our theoretical arguments and provide
insights into the detailed, time-evolving global wind and temperature
patterns and dynamical timescales (Section~\ref{model}).  We then
consider the dynamics of the stably stratified atmosphere that
overlies the convective interior, demonstrating how large-scale
vortices and/or zonal\footnote{Zonal refers to the east-west 
(longitudinal) direction whereas meridional refers to the north-south
(latitudinal) direction.  Zonal and meridional winds are winds
in the eastward and northward directions, respectively.  Zonal
jets refer to atmospheric jet streams oriented in the east-west direction.} jets are likely to emerge from interactions with
the interior (Section~\ref{atmosphere}).  Next, we 
consider observational implications, since IR radiation to space
typically occurs from within this stratified layer and hence infrared
spectra and lightcurves are strongly shaped by its dynamics
(Section~\ref{observables}).  The
final section concludes (Section~\ref{conclusions}).

We emphasize that our goal is to provide
a theoretical foundation for understanding the atmospheric dynamics 
of rapidly rotating, ultracool dwarfs and young EGPs, broadly
defined.  As such, we emphasize dynamical considerations, and
intentionally simplify our models by excluding clouds, chemistry,
and detailed representation of radiative transfer.   
This provides a clean environment
in which to identify key dynamical processes and construct
a theoretical foundation for more realistic studies
that will surely follow.  

\section{Background theory: Application to brown dwarfs}
\label{theory}

Here, we review basic concepts in atmospheric fluid dynamics
and apply them to brown dwarfs to understand the large-scale
structure of the flow.

\subsection{Importance of rotation}

Brown dwarfs rotate rapidly, and this will exert a major influence on
their atmospheric dynamics.  To demonstrate, consider the momentum
equation for a rotating fluid, which is given in the rotating
reference frame of the brown dwarf by
\begin{eqnarray}
\frac{d{\bf v}}{dt}+2\mathbf{\Omega}\times\mathbf{v} & = & 
-\frac{1}{\rho}\nabla p-\nabla\Phi,\label{eq:mom eq-1}
\end{eqnarray}
where $d{\bf v}/dt=\partial {\bf v}/\partial t + {\bf v}
\cdot\nabla {\bf v}$ is the material derivative, ${\bf v}$ is
the 3D velocity vector, $\mathbf{\Omega}$ is the planetary rotation vector,
$\Phi$ is a force potential which includes both the gravitational and 
centrifugal accelerations, $\nabla$ is the three-dimensional gradient
operator, $t$ is time, and $p$ and $\rho$ are the pressure and
density fields respectively \citep{pedlosky-1987}. For the purpose of
this discussion we will assume the dynamics are inviscid.  We also
for the present neglect the Lorentz force, as appropriate for the
atmospheres and molecular envelopes of cool brown dwarfs; nevertheless,
the Lorentz force will be crucial in the deep interior, and we will return
to a discussion of it in Section~\ref{convective-flow-organization}.

The nature of the flow depends on the Rossby number, given by the
ratio of the advective and Coriolis accelerations, $Ro = U/\Omega L$,
where $U$ is a characteristic wind speed, $L$ is a characteristic
length scale of the flow, and $\Omega$ is the rotation rate ($2\pi$
over the rotation period).  If $Ro\ll1$, the flow is rotationally
dominated; if $Ro\sim1$, rotation is important but not dominant, while
if $Ro \gg 1$, rotation plays a minor role \citep[see, e.g.,][p.~84]
{vallis-2006}.  Because of their fast rotation rates, Jupiter, Saturn,
Uranus, and Neptune, as well as the extratropical atmospheres of Earth
and Mars, all exhibit $Ro \ll 1$ at large scales; the only
solar-system atmospheres where rotation does not dominate are those of Titan
and Venus.

\begin{figure}
\includegraphics[scale=0.62, angle=0]{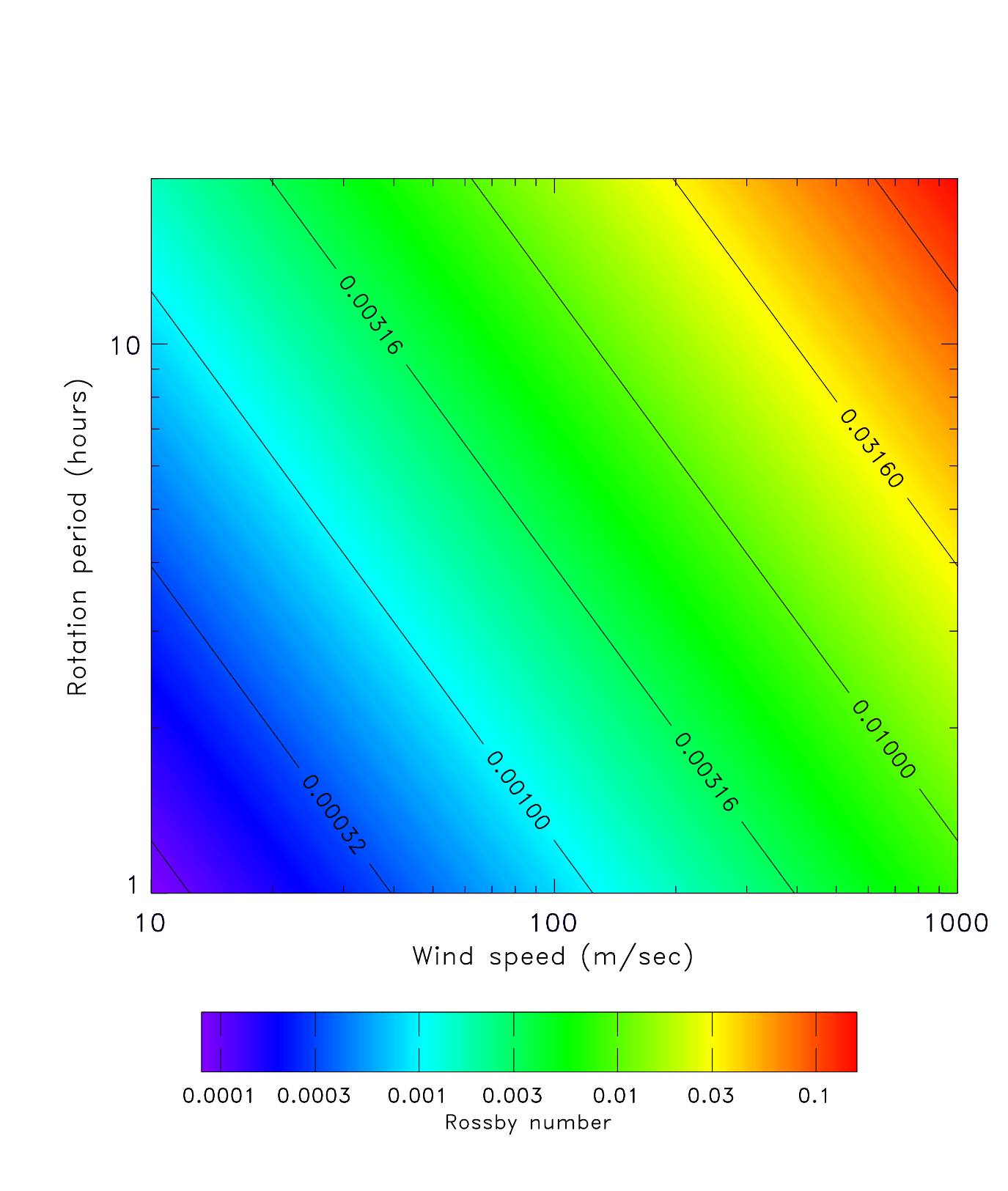}
\caption{Rossby numbers expected on brown dwarfs as a function of
  rotation period and characteristic wind speed, assuming dynamical
  features are global in scale (length equals one Jupiter radius).
  Rossby numbers range from $\sim$$10^{-4}$ to $0.1$, indicating that
  the regional- and global-scale dynamics in brown-dwarf atmospheres
  will be rotationally dominated over a wide range of parameters.
  Contour levels are in half-decade increments from 0.1 at the upper 
  right to 0.0001 at the lower left.}
\label{rossby}
\end{figure}

Estimating $Ro$ for brown dwarfs requires knowledge of wind speeds and
flow length scales, which are unknown.\footnote{Here, we seek to
  understand the global-scale flow and the appropriate values are
  therefore not the convective velocities and length scales but the
  wind speeds and length scales associated with any organized jets
  (i.e. zonal flows) and vortices that may exist.}  Nevertheless,
\citet{artigau-etal-2009} show that if the evolution of their light
curve shapes over intervals of days is interpreted as differential
zonal advection of quasi-static features by a latitude-dependent zonal
wind, the implied differential rotation is $\sim$1\%, which for the
2.4-hour rotation period of SIMP0136 implies a zonal wind speed of
$\sim$300--$500\rm\,m\,s^{-1}$ depending on the latitude of the
features.  A similar analysis by \citet{radigan-etal-2012} suggests a
possible zonal wind speed of $\sim$$45\rm\,m\,s^{-1}$ for 2M2139,
although they caution that this estimate relies on rather tentative
assumptions.  Interestingly, these values bracket the range of wind
speeds measured for the giant planets in the solar system, which range
from typical speeds of $\sim$$30\rm\,m\,s^{-1}$ on Jupiter to
$\sim$$300\rm\,m\,s^{-1}$ on Neptune \citep[e.g.,][]
{ingersoll-1990}.\footnote{The {\it maximum} observed speeds,
  expressed as a difference between the peak eastward and peak
  westward zonal winds, are several times these typical values,
  reaching $\sim$$200\rm\,m\,s^{-1}$ on Jupiter and
  $\sim$$600\rm\,m\,s^{-1}$ on Neptune.}  Later, we show that the
large-scale winds in the convective interior of a brown dwarf are
likely to be weak (Section~\ref{convective-flow-organization}), 
but that winds potentially exceeding $\sim$$10^2\rm\,m\,s^{-1}$
could develop in the stratified atmosphere (Section~\ref{atmosphere}).

Regarding length scale, the fact that SIMP0136
and 2M2139 exhibit large-amplitude variability hints that
atmospheric features could be near-global in size (particularly for
2M2139 where variability reaches 25\%).  This would imply $L\sim R_J
\sim 7\times10^7\rm\,m$, where $R_J$ is Jupiter's radius. On the other hand,
only a small fraction of brown dwarfs exhibit such large variability
and it is possible that length scales are typically smaller; for 
example, $L\sim10^7\rm\,m$ on Jupiter, Saturn, Uranus, and Neptune.

Adopting length scales $L\sim10^7$--$10^8\rm\,m$, wind speeds
$U\sim10$--$1000\rm\,m\,s^{-1}$, and rotation rates of 2 to 10 hours
yields Rossby numbers ranging from 0.0001 to 0.4.  Figure~\ref{rossby}
shows the Rossby number as a function of wind speed and rotation period
for the case of global-scale flows.  The values are much less than one
everywhere except for the largest wind speeds and slowest
rotation periods considered.  This implies that,
in general, the large-scale circulation on brown dwarfs will be
rotationally dominated.  The $Ro\ll 1$ condition on brown dwarfs
implies that the flow is geostrophically balanced,
that is, the primary balance in the momentum equation is between
Coriolis and pressure-gradient forces \citep{pedlosky-1987}.

\subsection{Organization of flow}

Significant insight into the flow structure can be obtained
from the vorticity balance.  Taking the
curl of \eqref{eq:mom eq-1} gives a vorticity equation of the form
\citep[e.g.,][]{pedlosky-1987}
\begin{equation}
\frac{\partial\omega}{\partial t}+\left(2\mathbf{\Omega}+\omega\right)\cdot\nabla\mathbf{v}-\left(2\mathbf{\Omega}+\omega\right)\nabla\cdot\mathbf{v}  =  -\frac{\nabla\rho\times\nabla p}{\rho^{2}},
\label{eq:vorticity-eq}
\end{equation}
where $\mathbf{\omega}=\nabla\times\mathbf{v}$ is the relative
vorticity.  The term on the right side, called the baroclinic term, is
nonzero when density varies on constant-pressure surfaces.  Note that,
since $\omega$ scales as $U/L$, the ratio $\omega/\Omega \sim Ro$, and
the time-derivative term is generally order $Ro$ smaller in magnitude
than the second and third terms on the left side.  Taking $Ro\ll1$,
appropriate to the flow on a brown dwarf, yields a leading-order
vorticity balance given by
\begin{eqnarray}
2\Omega\cdot\nabla\mathbf{v}-2\Omega\nabla\cdot\mathbf{v} & = & 
-\frac{\nabla\rho\times\nabla p}{\rho^{2}}.\label{eq:vorticity-eq-2}
\end{eqnarray}
One might expect that convection homogenizes the
entropy within the convection zone, in which case density does not
vary on isobars and $\nabla\rho\times
\nabla p=0$.  This is called a barotropic flow. In this case, 
Eq.~(\ref{eq:vorticity-eq-2}) simply becomes the compressible-fluid
generalization of the Taylor-Proudman theorem, which, expressed 
in a cylindrical coordinate system centered on the
rotation axis, is
\begin{equation}
\frac{\partial u}{\partial \hat z}=\frac{\partial v_\perp}{\partial \hat z}= 0
\label{tp1}
\end{equation}
\begin{equation}
\nabla_\perp\cdot {\bf v}_\perp = 0.
\label{tp2}
\end{equation}
where $\hat z$ is the direction parallel to the rotation axis, $u$ is 
the azimuthal (zonal) velocity, $v_\perp$ the velocity toward/away from the rotation
axis, ${\bf v}_\perp=(u,v_\perp)$ is the velocity in the plane 
perpendicular to the rotation axis, and $\nabla_\perp$ is the gradient 
operator in the plane perpendicular to the rotation axis.
The theorem states that if the flow has a small Rossby number and is inviscid
and barotropic the fluid motion will be completely two-dimensional,
therefore there will be no variation in the fluid velocity along the
direction of the axis of rotation.  The fluid then moves in columns
aligned with the rotation axis.  Note that no constant-density
assumption was made; Equations~(\ref{tp1})--(\ref{tp2})
hold in a barotropic, geostrophic, low-viscosity fluid even if
the density varies by orders of magnitude across the system.

Within a spherical planet or brown dwarf, such columns can easily move
in the zonal (east-west) direction.  However, the columns cannot
easily move toward or away from the rotation axis, because this
changes the length of the columns and the local density within them,
both of which induce non-zero $\nabla_\perp\cdot {\bf v}_\perp$ that
violate the theorem.  For such a barotropic fluid, the predominant
planetary-scale circulation therefore consists of zonal (east-west)
wind whose speed varies minimally in the direction along the rotation
axis; by comparison, the north-south flow is weak.  Of course, the
theorem is only valid to order $Ro$, and so motions toward/away from
the rotation axis---as well as motions parallel to the fluid
columns---can occur but only with amplitudes $\sim$$Ro$ less than that
of the primary zonal flow.\footnote{Of course, at very small scales,
  the Rossby number exceeds unity and the convection at these small
  scales will not organize into columns
  \citep[e.g.,][]{glatzmaier-etal-2009}.  The columnar organization
  applies only at scales sufficiently large that $Ro\ll 1$.}

In reality, turbulent convection results
in horizontal entropy gradients and therefore the fluid is not in
a barotropic state leading to a non-vanishing term on the right side
of Eq.~\eqref{eq:vorticity-eq-2}.  As a result, shear can develop
along the $\hat z$ direction.  Considering the zonal component
of Eq.~\eqref{eq:vorticity-eq-2} yields
\begin{equation}
2\Omega {\partial u\over\partial \hat z} = -{\nabla\rho\times\nabla p
\over\rho^2}\cdot \hat\lambda
\end{equation}
where $\hat\lambda$ is the unit vector in the longitudinal
direction.  If the flow exhibits minimal variation in longitude,
then it can be shown that $|\nabla\rho\times\nabla p|
= |\nabla p|(\partial\rho/\partial y)_p$.  Since $|\nabla p|$
is overwhelmingly dominated by the hydrostatic component, we have
to good approximation \citep{showman-etal-2010}
\begin{equation}
2\Omega {\partial u\over\partial \hat z}={g\over\rho r}
\left({\partial\rho\over\partial \phi}\right)_p,
\label{thermal-wind}
\end{equation}
where $\phi$ is latitude and $r$ is radial distance from the center
of the planet.  Thus, variations in the geostrophic wind along $\hat
z$ must be accompanied by variations in density on isobars.  This
relation, well-known in atmospheric dynamics, is called the
thermal-wind equation.

By itself, however, the preceding theory gives little insight into 
the spatial organization---columnar or not---of the internal entropy 
perturbations and any thermal-wind shear that accompanies them.
An alternative point of view that sheds light on this issue
is to consider the angular momentum budget.
The angular momentum per unit mass about the rotation axis is given by 
\begin{eqnarray}
M & \equiv & M_{\Omega}+M_{u}=\Omega r^{2}\cos^{2}\phi+ur\cos\phi,
\label{eq:am}
\end{eqnarray}
where the first and second terms represent the contributions due to
the planetary rotation and winds in that rotating frame, respectively.
Writing the zonal momentum equation in terms of angular momentum yields
\citep[e.g.,][Chapter 11]{peixoto-oort-1992}
\begin{equation}
\rho {dM\over dt} = -{\partial p\over\partial \lambda}.
\label{mom}
\end{equation}
It is useful to decompose the pressure and density into
contributions from a static, wind-free reference state and
the deviations from that state due to dynamics.  When wind 
speeds are much less than the speed of sound, these dynamical density
and pressure perturbations are small, leading to a continuity
equation $\nabla \cdot (\tilde\rho \mathbf{v})=0$, where
$\tilde\rho$ is the reference density profile.\footnote{The
reference density will generally be a function of radius.  Note that
this approximate continuity equation (essentially the anelastic
approximation) filters sound waves from the system, which is a reasonable
approximation as long as wind speeds are much less than the speed of sound.}
Motivated by the fact that the convective eddies drive a mean flow, 
we represent the dynamical variables as the sum of their
zonal means (denoted by overbars) and the deviations therefrom
(denoted by primes), such that $M=\overline{M} + M'$,
$\mathbf{v}=\overline{\mathbf{v}} + \mathbf{v'}$, etc.  Here,
we refer to these overbarred quantities as the mean flow and the
primed quantities as the eddies.  Substituting
these expressions into Eq.~(\ref{mom}) and zonal averaging
leads to the zonal-mean momentum equation \citep[cf][]{kaspi-etal-2009}
\begin{equation}
\tilde\rho{\partial \overline{M}\over\partial t} 
+ \nabla\cdot(\tilde\rho  \overline{\mathbf{v}} \,\overline{M})
+ \nabla\cdot(\tilde\rho \overline{\mathbf{v'} M'})=0.
\end{equation}
This equation states that temporal changes to the zonal-mean
angular momentum at any given location (first term) can only result from
advection of the zonal-mean angular momentum by the mean flow
(second term) or changes in the zonal-mean angular momentum due to
torques caused by eddy motions (third term).
In a statistical steady state, $\overline{M}$ equilibrates and
the equation simply becomes
\begin{equation}
 \nabla\cdot(\tilde\rho  \overline{\mathbf{v}} \,\overline{M})
+ \nabla\cdot(\tilde\rho \overline{\mathbf{v'} M'})=0.
\label{mom2}
\end{equation}
Since the ratio of $M_u$ to $M_\Omega$ is essentially the
Rossby number, it follows that for the $Ro\ll1$ regime expected
on a brown dwarf, $M\approx M_{\Omega}$.  Thus, for a rapidly rotating 
brown dwarf, surfaces of constant angular momentum are nearly parallel 
to the axis of rotation.  Using this result, along with the
continuity equation, Eq.~(\ref{mom2}) becomes
\begin{equation}
\overline{\mathbf{v}}\cdot\nabla {M_\Omega} =
-{1\over\tilde\rho}\nabla\cdot(\tilde\rho\overline{\mathbf{v'}M'}).
\label{mom3}
\end{equation}

This result has major implications for the circulation on rapidly
rotating giant planets and brown dwarfs. The equation implies that
the mean flow, $\overline{\mathbf{v}}$, can only cross angular momentum 
surfaces in the presence of eddy correlations between $\mathbf{v'}$ and 
$M'$ (such eddy correlations cause a torque that changes the
zonal-mean angular momentum following the mean flow, as necessary
for the mean flow to cross angular momentum surfaces).  When
such eddy effects are small, or if the flow is axisymmetric 
with no variation in longitude (for which $\mathbf{v'}$ and $M'$ are
zero by definition), then
$\overline{\mathbf{v}}\cdot\nabla M_{\Omega}=0$ \citep[][]{liu-schneider-2010}.  
In such a situation, the mean flow must to leading order be parallel to surfaces of
constant angular momentum \citep{busse-1976, kaspi-etal-2009} and
there can be no flow crossing these surfaces.  This constraint
places no limitation on the zonal-mean zonal flow, $\overline{u}$,
but requires the meridional circulation $\overline{v}$ to be small.
It is important to emphasize that this constraint differs from the 
Taylor-Proudman theorem, since it does not require the flow to be
barotropic, nor does it state that $\overline{u}$ is independent
of $\hat z$.

But how important are the eddy torques on the right-hand side of
Eq.~(\ref{mom3})?  At small Rossby number, the planetary rotation
contains so much angular momentum that even in the presence of
vigorous convection, eddy torques are unable to drive a rapid
mean-meridional circulation; therefore, we still have
$\overline{\mathbf{v}}\cdot\nabla M_\Omega \approx0$ at leading order.
To show this, we can estimate the timescale of the meridional
circulation and compare it to the characteristic timescale for
convection.  The characteristic timescale for convection to traverse
the interior is $\tau_{\rm conv} \sim D/w$, where $D$ is a
thickness of the layer in question (e.g., the planetary radius) and
$w$ is the characteristic convective speed.  We can
estimate the timescale for the mean flow to cross angular momentum
contours as follows.  The eddy correlation $\overline{\mathbf{v'}M'}$
is just $\overline{\mathbf{v'}u'r_\perp}$, where $r_\perp$ is the
distance from the rotation axis.  Under the assumption that the eddy
velocities scale with the convective velocities $w$, we can
write $\overline{\mathbf{v'}M'} \sim C r_\perp w^2$ where
$C$ is a correlation coefficient equal to one when $u'$ and
$\mathbf{v'}$ are perfectly correlated and equal to zero when $u'$ and
$\mathbf{v'}$ exhibit no correlation.  To order of magnitude,
Eq.~(\ref{mom3}) then becomes
\begin{equation}
\overline{v}_\perp\Omega r_\perp \sim C w^2
\end{equation}
and the meridional velocity therefore has a characteristic magnitude
\begin{equation}
\overline{v}_\perp\sim {C w^2\over \Omega r_\perp}
\end{equation}
Defining a timescale for the meridional circulation, $\tau_{\rm merid}
= r_\perp/\overline{v}_\perp$, implies that
\begin{equation}
\tau_{\rm merid} \sim \tau_{\rm conv} {r_\perp \Omega\over C w}
\end{equation}
which can be expressed as
\begin{equation}
\tau_{\rm merid} \sim {\tau_{\rm conv} \over  C \, Ro_{\rm conv}}
\end{equation}
where $Ro_{\rm conv} = w/r_\perp\Omega$ is a convective
Rossby number giving the ratio of the convective
velocities to the typical rotational velocity of the planet in
inertial space.  For typical brown dwarfs, where rotational 
velocities are tens of $\rm km\,s^{-1}$,
we expect $Ro_{\rm conv}\ll 1$; given the expected
convective velocities (see \S\ref{model}), we expect
$Ro_{\rm conv}\sim 10^{-5}$--$10^{-2}$.  Therefore,
the timescale for the meridional circulation is several
orders of magnitude longer than the characteristic convection
timescale.

This also means that the convective heat transport will be more
efficient along (rather than across) surfaces of constant angular
momentum. Fig.~\ref{convect} shows the onset of convection for an
experiment driven by a constant heat flux at the bottom boundary.  Two
models are shown, a rapidly rotating case on the left and a slowly
rotating case on the right.  While for the larger Rossby number case
the dominant driving force for the turbulent plumes is the buoyancy
and therefore the plumes are driven away from the center of gravity,
for the small Rossby number case the convection becomes aligned along
the direction of the axis of rotating demonstrating the angular
momentum constraint ($\mathbf{v}\cdot\nabla M_{\Omega}=0$).  Note that
for small Rossby number experiment only close to the boundaries does
the Rossby number approach one and therefore there the convective
cells can close. Thus rotation strongly modulates the heat transport
from the interior of the brown dwarf at large scales.

\section{Three-dimensional circulation model of convection-zone dynamics}
\label{model}

\subsection{Model}

We solve the fluid equations for a convecting, 
three-dimensional, rotating brown dwarf.   We adopt the
anelastic system \citep[e.g.,][]{ogura-phillips-1962, gilman-glatzmaier-1981,
ingersoll-pollard-1982}, which assumes that dynamics introduces only small
perturbations of the density, entropy, and pressure from a
specified reference state, which we here take to be isentropic.
Dynamical density perturbations then enter the
momentum equations in the buoyancy term but do not appear in the
continuity equation; this has the effect of filtering acoustic
waves from the system.
The anelastic system is appropriate for the fluid interior of a 
brown dwarf, where dynamical perturbations of entropy, density, and
pressure due to convection are expected to be modest and convection 
should lead to a nearly constant entropy throughout.  Although this
study represents its first application to brown
dwarfs, the anelastic system has previously been used with great success for
understanding convection in Jupiter and Saturn \citep{kaspi-etal-2009, 
jones-kuzanyan-2009, glatzmaier-etal-2009, showman-etal-2011} 
and stellar interiors \citep[and references therein]{miesch-toomre-2009}.

Our particular
implementation is that of \citet{kaspi-etal-2009}.  The momentum,
continuity, and energy equations, respectively, are given by
\begin{equation}
{\partial {\bf v}\over\partial t} + (2{\bf \Omega} + {\bf \omega})\times
{\bf v}=-{1\over\tilde\rho}\nabla p' - {\rho'\over\tilde\rho}\nabla\Phi
- {1\over2}\nabla{\bf v}^2 + \nu \nabla^2{\bf v}
\label{momentum}
\end{equation}
\begin{equation}
\nabla\cdot(\tilde\rho {\bf v})=0
\label{continuity}
\end{equation}
\begin{equation}
{\partial s'\over\partial t} + {1\over\tilde\rho}\nabla\cdot
(\tilde\rho {\bf v}s') - {1\over\tilde\rho}\nabla\cdot(\tilde\rho
{\bf \kappa}\nabla s') = {Q\over \tilde T}
\label{energy}
\end{equation}
where $Q$ is thermodynamic heating/cooling per mass, 
$\nu$ is the kinematic viscosity, $\kappa$ is the thermal diffusivity,
and other quantities are as defined previously.
Here, both $\nu$ and $\kappa$ are taken as constants and are intended
to parameterize small-scale eddy mixing.
The quantities $\tilde\rho(r)$, $\tilde p(r)$, and $\tilde T(r)$
are the radially varying reference profiles of density, pressure, 
and temperature, respectively; $\rho'$ and $p'$ are the deviations 
of the density and pressure from their local reference values,
such that the total pressure and density are $\rho=\tilde\rho + \rho'$
and $p=\tilde p + p'$.  Likewise, $s'$ is the deviation of entropy
from its reference state value.

The system is closed with an equation of state, which 
enters through the reference profiles and through the relationship
between the density, entropy, and pressure perturbations in the
anelastic system
\begin{equation}
{\rho'\over\tilde\rho}={1\over\tilde\rho}\left({\partial\rho\over
\partial s}\right)_p s' + {1\over\tilde\rho}\left({\partial\rho
\over\partial p}\right)_s p' \equiv -\alpha_s s' + \beta p',
\end{equation}
where $\alpha_s$ and $\beta$ are the radially varying isobaric entropy
expansion coefficient and isentropic compressibility, respectively,
along the model's radially varying reference profile.  Here, we adopt
the SCVH equation of state (EOS) for hydrogen-helium mixtures
\citep{saumon-etal-1995}.  Given a specified brown-dwarf mass and
internal entropy, and the assumption that the reference state is in
hydrostatic balance\footnote{We emphasize that the dynamical model
  itself is non-hydrostatic; hydrostatic balance is used only in
  defining the reference state.}, this EOS allows us to calculate the
radially varying reference profiles $\tilde\rho$, $\tilde p$, and
$\tilde T$ \citep[e.g.,][]{guillot-morel-1995, guillot-etal-2004} as
well as the radial profiles of $\alpha_s$ and $\beta$ along the
reference adiabat.  The gravitational acceleration in the model
varies radially, which we determine by integrating this basic state
radially.  See \citet[][Figure 2]{kaspi-etal-2009} for the resulting
radial profiles of density, temperature, pressure, gravity, thermal
expansion coefficient, and specific heat used in the model.

Many studies of convection in rotating spherical shells force the
system by passing a heat flux through impermeable upper and lower
boundaries, either with a constant-temperature or constant heat flux
boundary condition \citep[e.g.,][and many others] {christensen-2001,
  christensen-2002, aurnou-olson-2001, heimpel-etal-2005}.  However,
this is unrealistic in the context of a substellar object.  At high
Rayleigh numbers, passing a heat flux through the model boundaries
will lead to thin hot and cold boundary layers at the bottom and top
boundaries, respectively, which detach and form hot and cold
convective plumes that in some cases can dominate the dynamics.
Because real brown dwarfs are fluid throughout, the bottom boundary
layer, in particular, is unrealistic.  Instead, we force the system by
imposing a vertically distributed source of internal heating and
cooling throughout the bottom and top portions of the domain,
respectively, thus allowing outward convective transport of heat
without the development of artificial boundary layers \citep[for more
  detail see][] {kaspi-etal-2009}.\footnote{Real brown dwarfs of
  course do not have substantial internal heat sources (the burning of
  deuterium not playing a role except in the most massive objects) but
  rather decrease in entropy very gradually over multi-Byr timescales.
  Thus, ideally, one would like to set up the problem with a heat sink
  near the top (i.e., cooling) and no heat source near the bottom,
  thereby allowing the internal entropy to decline with time in a
  brown-dwarf like fashion.  The difficulty is that, due to
  computational limitations, achieving steady state requires the
  system to be overforced \citep{showman-etal-2011}, and without a
  source of energy near the bottom the interior entropy would decline
  unrealistically rapidly.  Adding a heat source near the bottom, as
  we have done, allows the global-mean interior entropy to be
  essentially constant over dynamical timescales, consistent with
  expectations for brown dwarfs.}  The top and bottom thermal boundary
conditions correspond to zero heat flux.  The top and bottom
mechanical boundary conditions are impermeable in radial velocity and
free-slip in horizontal velocity.

We solve the equations in spherical geometry using the
state-of-the-art circulation model MITgcm \citep{adcroft-etal-2004},
which \citet{kaspi-2008} adapted for anelastic simulations of the deep
convective envelopes of giant planets.  The equations are solved using
a finite-volume discretization on a staggered Arakawa C grid
\citep{arakawa-lamb-1977} in longitude and latitude.  Our typical
resolution is $1^{\circ}$ in longitude and latitude with 120 vertical
levels spaced to give enhanced resolution near the top of the domain
where the pressure and density scale heights are the smallest
\citep[see][]{kaspi-etal-2009}.  Most models extend the full
$360^{\circ}$ in longitude and in latitude from $80^{\circ}$S to
$80^{\circ}$N.  For some parameter variations, we performed
simulations in sectors $90^{\circ}$ of longitude wide (using a
periodic boundary condition between the eastern and western
boundaries) with a resolution of $2^{\circ}$ of longitude and latitude
and 120 vertical levels.  For all models, the outer and inner
boundaries are spherical surfaces with radii of $1 R_J$ and $0.5 R_J$
respectively.  This choice of inner boundary is sufficiently deep to
minimize any artificial effect of the lower boundary on the surface
dynamics.  We generally use a Jupiter-like interior reference profile
with a pressure at the outer surface of 1 bar and 20 Mbar at the
bottom boundary.  The interior reference density varies by a factor of
over $10^4$ from the 1-bar level to the deep interior
\citep[see][]{kaspi-etal-2009}.  All simulations are spun up from rest
using an initial thermal profile corresponding to the reference
profile, and are integrated until a statistical steady state is
achieved.

\subsection{Results: Convective and thermal structure}

We perform simulations using rotation periods
ranging from 3 to 200 hours (spanning the typical range observed
for brown dwarfs) as well as an additional sequence of parameter variations 
adopting rotation periods as long as 2000 hours to illustrate the effect 
of rotation on the dynamics.

\begin{figure}
\includegraphics[scale=0.47, angle=0]{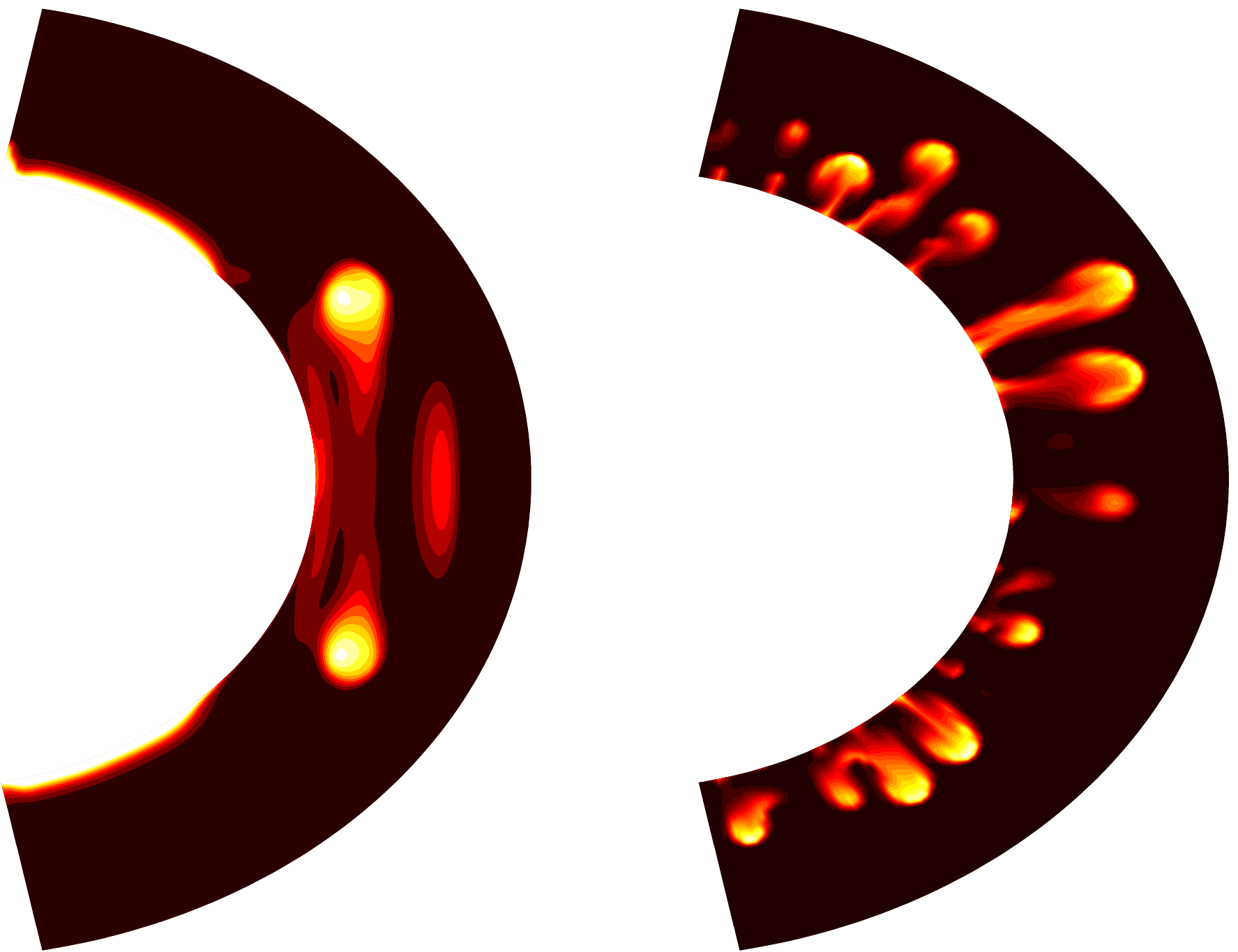}\\
\includegraphics[scale=0.47, angle=0]{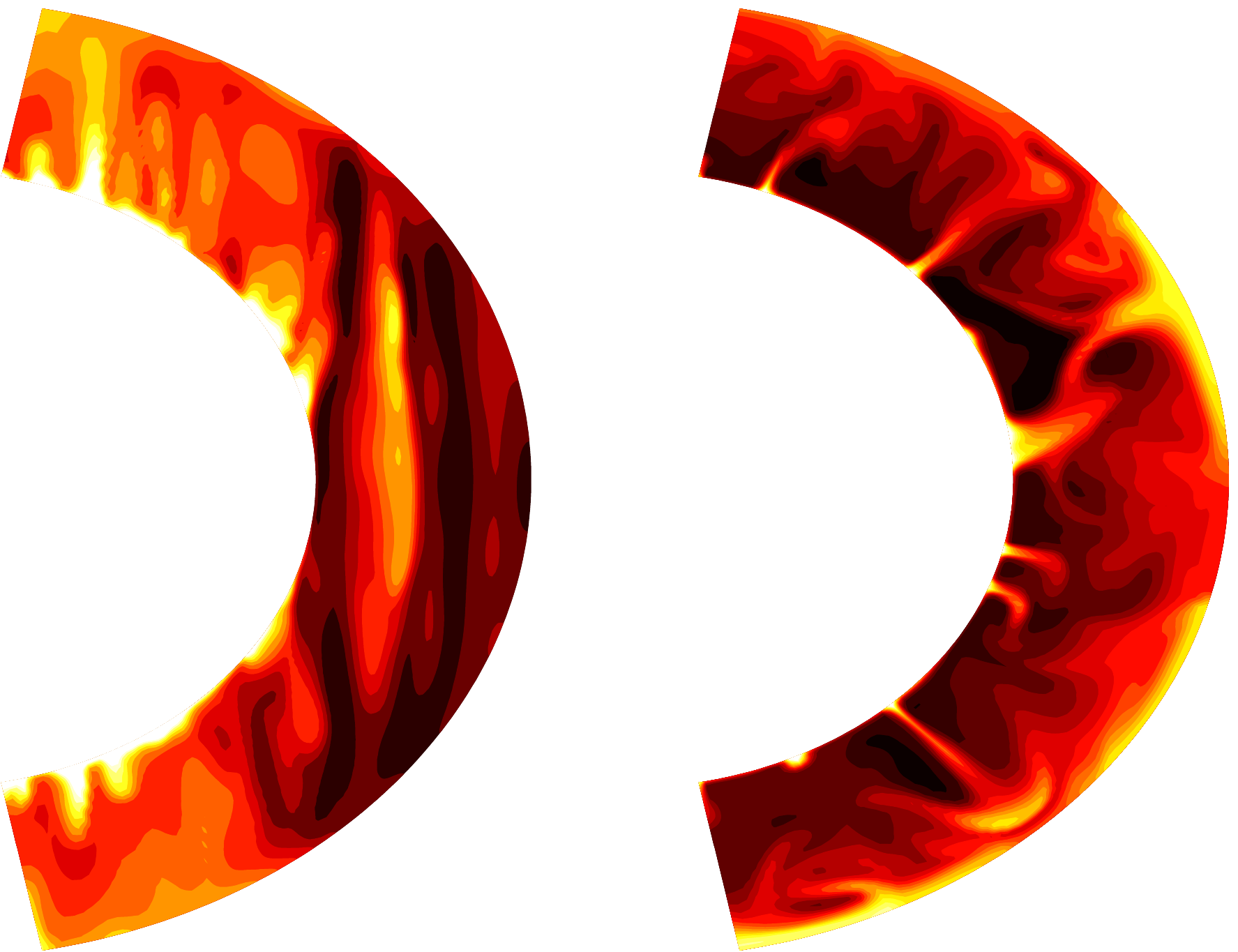}
\caption{A pedagogical illustration, using two anelastic models, of
  the importance of rotation in the brown-dwarf parameter regime.  The
  left column shows a rapidly rotating model (10 hour rotation
  period), and the right column shows a slowly rotating model (2000
  hour rotation period).  A constant heat flux is applied at the
  bottom boundary, leading to convection.  Both models are fully
  three-dimensional simulations extending $360^{\circ}$ in longitude
  and adopt Jovian-like radial profiles of density and thermal
  expansivity from the SCVH equation of state.  In each model, the top
  panel shows the transient initial stage soon after convection
  initiates, and the bottom panel shows the state after the convection
  is well developed.  Colorscale denotes entropy perturbations at an
  arbitrary longitude, shown in the radius-latitude plane.  Rotation
  vector points upward in the figure.  In the slowly rotating case,
  rotation plays no role in the dynamics, whereas in the rapidly
  rotating case, the rotation forces the large-scale flow to align
  along columns parallel to the rotation axis.}
\label{convect}
\end{figure}

Before presenting models for fully equilibrated brown dwarfs, we first
demonstrate with a pedagogical example the crucial importance that
rotation plays in the brown-dwarf parameter regime.
Figure~\ref{convect} depicts the temperature structure during the
spin-up phase for two models that are identical except for the
rotation period, which is 2000 hours in the model on the right and 10
hours for the model on the left.  The models in Figure~\ref{convect}
are not intended to be realistic brown dwarf models (for example, they
are forced by a heat flux from the bottom boundary, which is not
realistic in the context of a brown dwarfs) but are instead simply an
illustration of the importance of rotation in the brown dwarf
parameter regime.  Nevertheless, the models do have a realistic
Jovian interior structure, with density increasing by a factor of 
$\sim$$10^4$ from the interior to the exterior.

In the slowly rotating model (right panels of Figure~\ref{convect}),
the Coriolis forces are sufficiently weak that the Rossby number is
$\gtrsim$1, so that rotation plays a negligible role in the dynamics.
Convective plumes rise from the lower boundary and ascend
quasi-radially toward the outer boundary.  The plumes are equally able
to radially traverse the domain whether they emanate from the polar or
equatorial regions, and to zeroth order, the convection appears to be
isotropic.

In contrast, the rapidly rotating model exhibits \break $Ro\ll1$ and is thus
rotationally dominated (Fig.~\ref{convect}, left panels).  As predicted by
the theory in Section~\ref{theory}, the convection develops a columnar
structure.  Plumes that emerge in the polar regions can ascend and
descend radially while remaining at nearly constant distance from the
rotation axis; they therefore easily traverse the domain.  But plumes
forming at lower latitudes cannot easily cross the domain
because doing so requires them to change distance significantly from
the rotation axis.  In a $Ro\ll 1$ flow, angular momentum is not
homogenized and lines of constant angular momentum are nearly parallel
to the rotation axis.  As a result, moving toward or away from the
rotation axis can only be achieved by significantly increasing the
angular momentum of ascending fluid parcels or decreasing the angular
momentum of descending fluid parcels.  The timescale for this angular
momentum exchange is longer than the typical convection timescale for
plumes to traverse the domain, and thus convection toward or away from
the rotation axis is less efficient.  Therefore, as expected from
Section~\ref{theory}, rotation imposes on the flow a columnar
structure.  

\begin{figure}
\includegraphics[scale=0.47, angle=0]{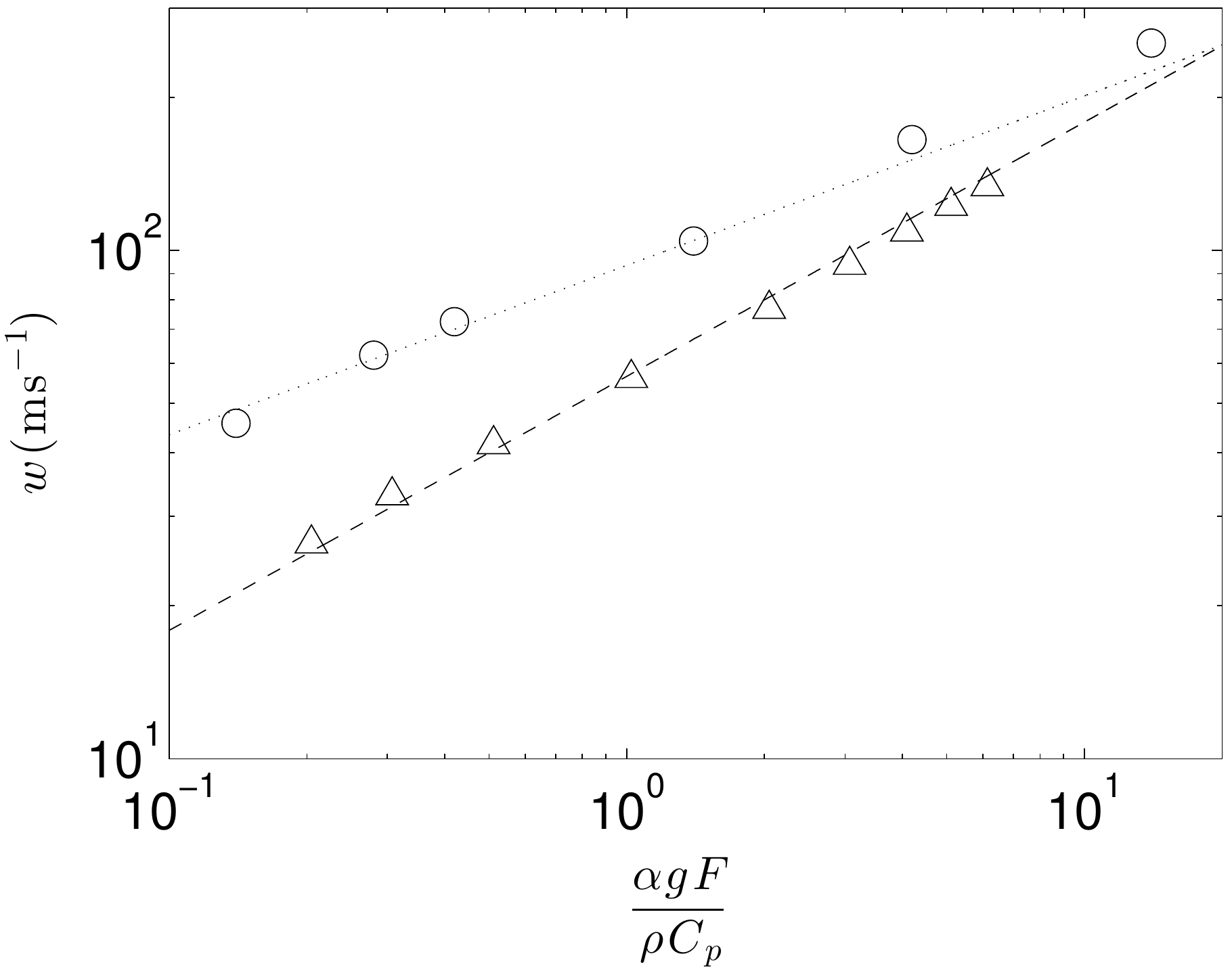}
\caption{Vertical (i.e., radial) velocities for brown dwarf convection
  models showing that rotation significantly affects the convective
  velocities.  Each symbol shows the domain-averaged, mass-weighted
  root-mean-square vertical wind speed versus the mass-weighted mean
  buoyancy flux $\alpha g F/\rho c_p$ (units $\rm m^2\,s^{-3}$) for a
  given numerical integration.  Circles show slowly rotating models
  (rotation period 100 hours), while triangles show rapidly rotating
  models (rotation period 10 hours).  For each rotation period, models
  with a range of buoyancy fluxes were performed.  The dotted and
  dashed lines show Equations~(\ref{w-nonrotate}) and
  (\ref{w-rotate}), respectively.}
\label{vert}
\end{figure}

Rotation strongly affects the vertical convective velocities as well.
To order of magnitude, convective velocities $w$ 
and temperature perturbations $\delta T$
relate to the convective heat flux $F$ as
\begin{equation}
F\sim \rho w c_p\delta T
\label{flux}
\end{equation}
where $c_p$ is specific heat at constant pressure and $\rho$
is the local density.  Convective temperature perturbations relate to 
convective density perturbations $\delta \rho$ via $\alpha \,\delta T 
\sim \delta \rho/\rho$, where $\alpha$ is thermal expansivity.  The
standard nonrotating mixing-length scaling results from 
assuming that buoyancy forces $g \,\delta\rho/\rho$
cause free acceleration of convective plumes over a mixing length $l$,
yielding \citep[e.g.,][]{clayton-1968, stevenson-1979}
\begin{equation}
w \sim \left({\alpha g F l\over \rho c_p}\right)^{1/3}.
\label{w-nonrotate}
\end{equation}
In contrast, in a rapidly rotating convective flow, convective
buoyancy forces often approximately balance vertical Coriolis
forces.  Assuming the turbulent motions are approximately isotropic
(i.e., horizontal eddy velocities are comparable to vertical
convective velocities), one instead obtains a vertical velocity
\begin{equation}
w \sim \gamma \left({\alpha g F\over \rho c_p \Omega}\right)^{1/2}.
\label{w-rotate}
\end{equation}
where we have introduced a dimensionless prefactor $\gamma$ that
is expected to be order unity.  Laboratory experiments in rotating
tanks demonstrate that this expression works well in explaining
the convective velocities in the rapidly rotating regime 
\citep{golitsyn-1980, golitsyn-1981, boubnov-golitsyn-1990,
fernando-etal-1991}.  A similar expression has also been
suggested for the dynamo-generating region of planetary interiors
where a three-way force balance between buoyancy, Coriolis,
and Lorentz forces may prevail \citep{starchenko-jones-2002,
stevenson-2003, stevenson-2010}.  \citet{showman-etal-2011} showed that
it also provides a good match for convective velocities
under Jupiter conditions.

\begin{figure}
\includegraphics[scale=0.52, angle=0]{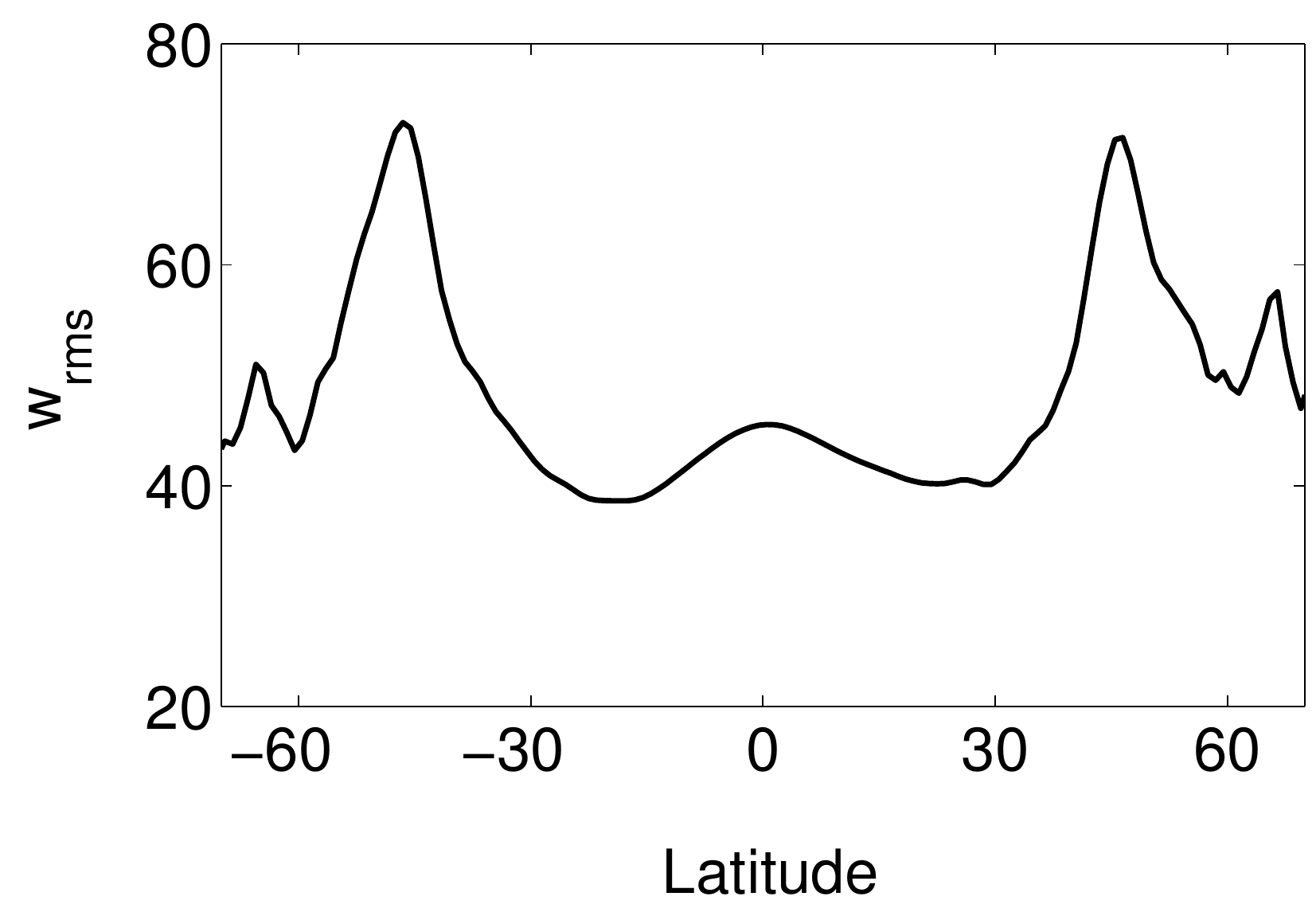}
\caption{Characteristic vertical convective velocities versus latitude
  in a convection model with a rotation period of 10 hours,
  indicating that convective velocities tend to be greater at
  mid-to-high latitudes than at low latitudes.
 Velocities are calculated at a given latitude and pressure
  as the root-mean-square in longitude and time. }
\label{w-latitude}
\end{figure}

Our models demonstrate that, under typical brown dwarf conditions, the
rotating scaling (\ref{w-rotate}) provides a significantly better
match to the convective velocities than the non-rotating scaling
(\ref{w-nonrotate}).  This is illustrated in Fig.~\ref{vert}, which
shows vertical velocities for our fully equilibrated brown dwarf
models.\footnote{These, and all subsequent, models are fully
  equilibrated brown dwarf models forced by internally distributed
  heating and cooling, thereby (unlike Figure~\ref{convect}) avoiding the
  generation of unrealistic lower thermal boundary layers.}  Symbols
depict the mass-weighted, global-mean vertical velocities for a
sequence of models that are slowly rotating (circles, rotation period
100 hours) and rapidly rotating (triangles, rotation period 10 hours).
They are plotted against the mass-weighted, global-mean buoyancy flux,
$\alpha g F/\rho c_p$, for each model.

There are several points to note in Figure~\ref{vert}.  First, the
velocities in the rapidly rotating models are smaller than in the
slowly rotating models, indicating the rotational suppression of
convective motions.  Second, the models show that the dependence of
vertical velocity on buoyancy flux is weaker in the slowly rotating
case than in the rapidly rotating case.  The dotted line shows the
nonrotating scaling (\ref{w-nonrotate}), with a mixing length of
$816\rm\,km$, while the dashed line gives the rotating scaling
(\ref{w-rotate}), with a prefactor $\gamma=0.75$.  The agreement is
good, showing that, in our slowly rotating models, the mass-weighted
mean vertical velocities scale approxiately as buoyancy flux to the
one-third power, whereas in our rapidly rotating models, the
mass-weighted mean vertical velocities scale approximately as buoyancy
flux to the one-half power---just as predicted by
Equations~(\ref{w-nonrotate}) and (\ref{w-rotate}), respectively.  At
a given buoyancy flux, the two scalings shown in Figure~\ref{vert}
differ by only a factor of a few, but the discrepancy becomes greater
with decreasing buoyancy flux, and the two predictions differ
significantly for conditions inside a typical L/T dwarf.  Adopting
parameters appropriate for an L/T dwarf ($\alpha\approx
10^{-5}\rm\,K$, $c_p \approx 10^4\rm J\,kg^{-1}\,K^{-1}$, and $F\sim
10^4$--$10^5\rm\,W\,m^{-2}$, corresponding to effective temperatures
of $\sim$$650$ to $1150\rm\,K$) yields buoyancy fluxes appropriate to
the bulk interior of $\sim$$10^{-6}$ to $10^{-5}\rm\,m^2\,s^{-3}$.
For these values, the convective velocities predicted by the rotating
scaling are an order of magnitude lower than those predicted by the
non-rotating scaling.  

Under conditions appropriate to a typical T
dwarf, Equation~(\ref{w-rotate}) predicts velocities of
$\sim$$0.1\rm\,m\,s^{-1}$ in the deep interior,
$\sim$$10\rm\,m\,s^{-1}$ at 1000 bar and $\sim$$40\rm\,m\,s^{-1}$ at
100 bar.  However, the equation likely overpredicts the velocities
near the top of the convection zone.  In particular, because
$\alpha/\rho$ is large near the outer boundary, the buoyancy forces
are large, and this likely implies a breakdown of
Equation~(\ref{w-rotate}) in the outermost part of the convection
zone.  Interactions of convection with the radiative-convective
boundary may also be important in modifying the convective velocities
there, an effect not included in Equation~(\ref{w-rotate}).

The convective velocities tend to be greater at high latitudes
than at low latitudes, as expected from angular momentum
constraints.  This is illustrated in Fig.~\ref{w-latitude} for
a model with a rapid (10 hour) rotation period.  Radial convective
motion at low latitudes can only occur if fluid parcels gain or lose
significant angular momentum as they change distance from the
rotation axis; in contrast, convective motion near the poles involves
comparatively modest changes in distance from the rotation axis and can occur 
more readily.  The result is greater convective velocities near the
poles than the equator. 
Nevertheless, rotational constraints still influence high-latitude convection:
the continuity equation demands that the vertical 
convective motion must necessarily be accompanied by horizontal convergence 
and divergence and thus motions toward or away from the rotation axis.
At large scales, when the Rossby number is small,
such rotational constraints will still play an inhibiting role in
the efficiency of polar convection.  This may explain why the velocities
in Fig.~\ref{w-latitude} vary by only a factor of $\sim$2 from equator
to pole.

\begin{figure}
\includegraphics[scale=0.6, angle=0]{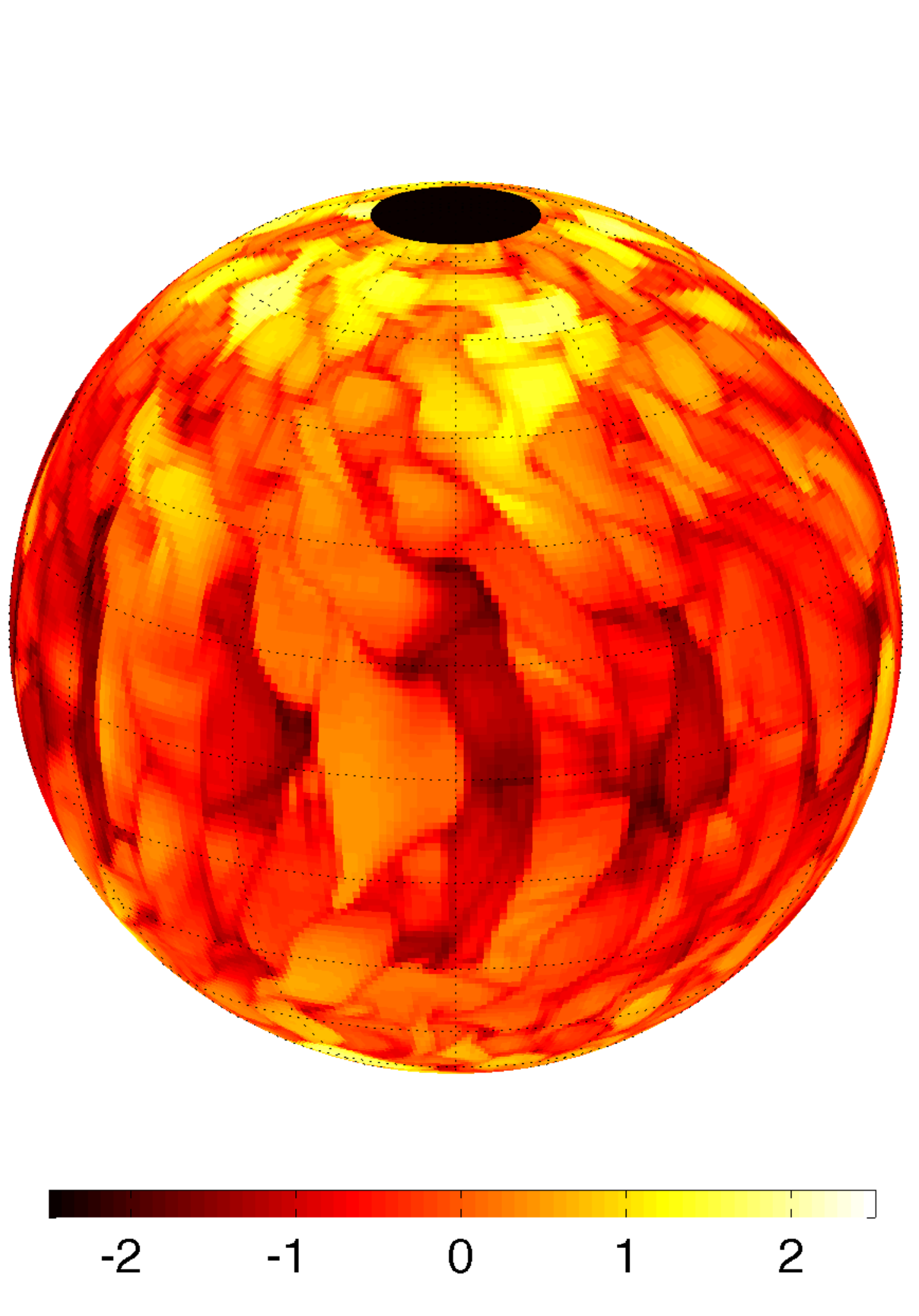}
\caption{Temperature variations at 1 bar in a typical, rapidly
rotating brown dwarf model.  Colorscale gives the temperature
perturbation $T'$ at 1 bar, that is, the deviation of temperature
from its reference value.  Rotation period is 10 hours.}
\label{globe-temp}
\end{figure}

We now examine the large-scale flow in our fully equilibrated brown
dwarf models.  The temperatures develop significant spatial structure
at regional-to-global scales, as shown at the 1-bar level in
Fig.~\ref{globe-temp} for a typical model.  Over a wide range of
conditions, the temperatures exhibit significant latitudinal
gradients, with polar temperatures exceeding equatorial temperatures
by typically a few K.  This equator-to-pole temperature difference
results from the greater efficiency of convection in polar regions
than equatorial regions.  Cooling to space continually decreases the
entropy of fluid near the $\sim$$1\,$bar level; in polar regions, this
low-entropy fluid readily sinks and is replaced with higher-entropy
material rising from below.  But in equatorial regions, the
suppression of radial convection inhibits the dense, low-entropy fluid
at $\sim$$1\,$bar from readily sinking.  The characteristic
hot-poles-cold-equator pattern seen in Fig.~\ref{globe-temp} is the
result.  We emphasize that this effect emerges naturally from the
dynamics and is not the result of any latitudinally varying forcing;
indeed, our forcing and boundary conditions are independent of
latitude.  

The temperature patterns also develop significant variations in both
longitude and latitude on regional scales of typically
$\sim$$10^7\rm\,m$ (Fig.~\ref{globe-temp}).  Convection produces
regional-scale thermal anomalies that vary substantially in time.  At
high latitudes, these regional anomalies tend to exhibit comparable
longitudinal and latitudinal scales, as might be expected from the
fact that the plumes move nearly vertically and converge or diverge
horizontally in a quasi-isotropic fashion there.  At low latitudes,
however, the structures exhibit north-south elongation.  This is the
manifestation of the columnar structure taken by the convection at
relatively large scales.  Note that real brown dwarfs also likely
exhibit short-lived convective structure  at very
small scales (e.g., granulation) that would be superposed on the larger-scale structure
like that shown in Figure~\ref{globe-temp}.  Resolving such small-scale structure
in global models would require simulations at significantly higher spatial
resolution than explored here, which will be a computational challenge
for the future.

\begin{figure}
\includegraphics[scale=0.52, angle=0]{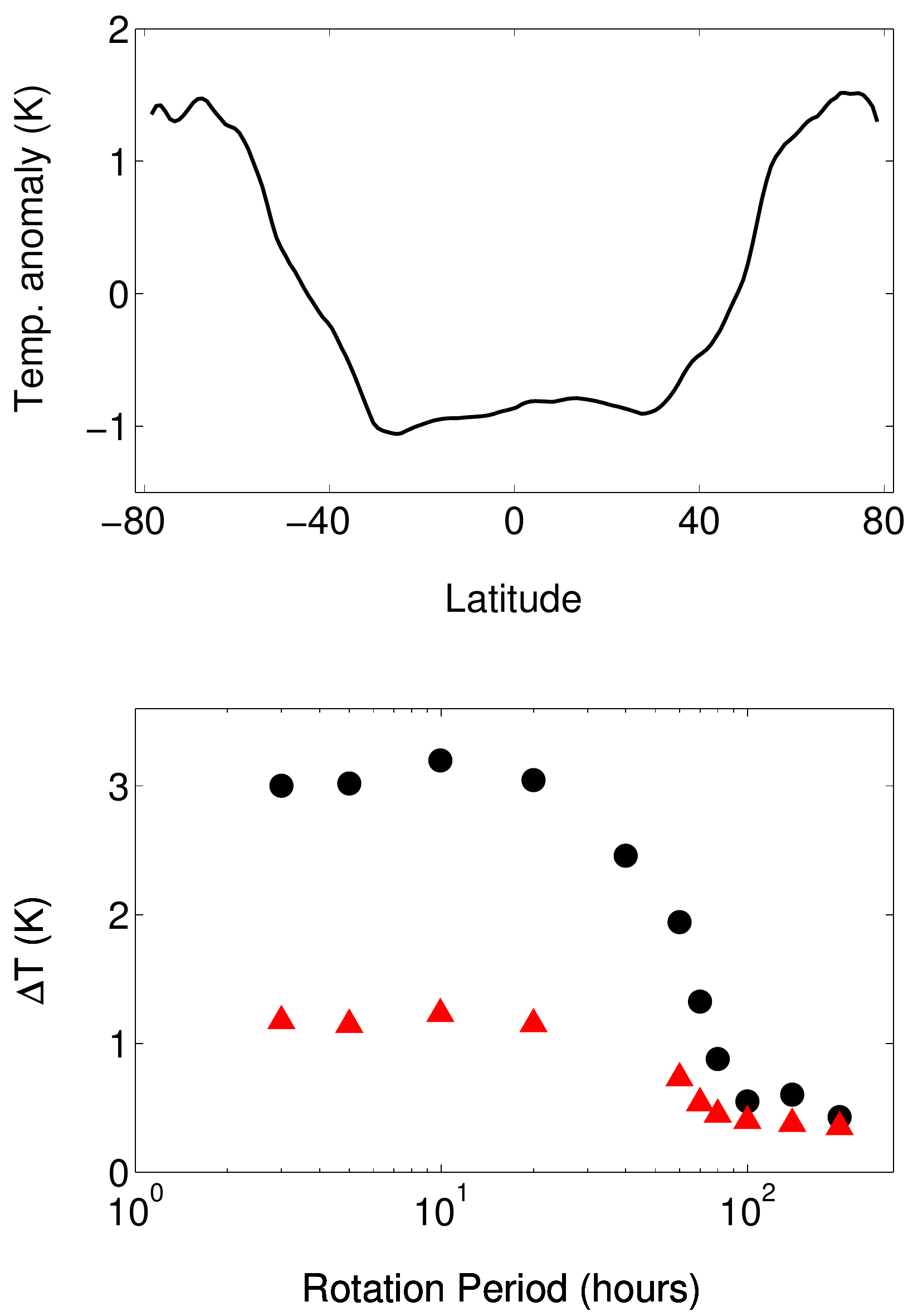}
\caption{{\it Top:} Longitudinal-mean temperature versus latitude at 1
  bar for a model with a rotation period of 10 hours, illustrating the
  emergence of a systematic equator-to-pole temperature difference.
  {\it Bottom:} Black dots show longitudinally averaged
  pole-to-equator temperature differences, and red triangles denote
  the root-mean-square horizontal temperature fluctuations, both at 1
  bar, for a sequence of otherwise identical models varying in
  rotation period from 3 to 200 hours. }
\label{T-vs-lat}
\end{figure}

The characteristic convective temperature perturbations
and equator-to-pole temperature differences in our models decrease with
increasing rotation period.  This is illustrated in Figure~\ref{T-vs-lat}.
The top panel shows the longitudinal (zonal) mean temperature
versus latitude at the 1-bar level for a model with a rotation
period of 10 hours, illustrating the hot poles and cold equator
with a difference of $\sim$$2\rm\,K$.  The bottom panel shows
the equator-to-pole temperature difference (black circles),
and root-mean-square (rms) temperature variations (red triangles), both
at the 1-bar level, for a sequence of models with differing rotation periods.
Both the equator-to-pole temperature differences and rms temperature
perturbations are nearly constant from rotation periods of
3 hours to $\sim$40 hours.  At rotation periods exceeding $\sim$50 hours,
however, the temperature perturbations decrease significantly.
This results from the fact that, at long rotation periods, the
Rossby number becomes large and the convection is no longer rotationally
inhibited.

The temperature contrasts expected in the convecting region
can be understood by combining Eqs.~(\ref{flux}) and (\ref{w-rotate})
to yield a relation for the convective temperature perturbations 
in a rotationally dominated flow \citep[cf][]{showman-etal-2010}
\begin{equation}
\delta T \sim \left({F\Omega\over \rho c_p \alpha g}\right)^{1/2}.
\label{deltaT-rotate}
\end{equation}
Our models are performed for Jovian-like internal profiles,
corresponding to $\Omega=1.74\times10^{-4}\rm\,s^{-1}$,
$c_p=1.3\times10^4 \rm\,J\,kg^{-1}\,K^{-1}$, and gravity, density, and
thermal expansivity at the 1-bar level of $23\rm\,m\,s^{-2}$,
$0.2\rm\,kg\,m^{-3}$, and $0.006\rm\,K^{-1}$, respectively.  As
discussed in detail by \citet{showman-etal-2011}, global convective
models of giant planets must, for computational reasons, be overforced
by several orders of magnitude; our model adopts a heat flux near
1-bar that is close to $10^7\rm\,W\,m^{-2}$.  For these values,
Eq.~(\ref{deltaT-rotate}) predicts $\delta T\sim 2\rm\,K$, very
similar to the values actually occurring in our models (e.g.,
Fig.~\ref{globe-temp}).  This indicates that Eq.~(\ref{deltaT-rotate})
provides a reasonable representation of the model behavior.
Extrapolating now to the conditions of a typical L/T dwarf, we adopt a
temperature of $1000\rm\,K$, corresponding to a radiated IR flux of
$F\sim 6\times10^4\rm\,W\,m^{-2}$.  Inserting parameters appropriate
to the 1-bar level of a brown dwarf ($\rho=0.03\rm\,kg\,m^{-2}$, $c_p
= 1.3\times10^4\rm\,J\,kg^{-1}\,K^{-1}$, $\alpha = 10^{-3}\rm\,K$,
$\Omega = 3\times10^{-4}\rm\,s^{-1}$, and $g=200\rm\,m\,s^{-2}$), we
obtain $\delta T\sim 0.5\rm\,K$ as the expected convective temperature
perturbation for a typical brown dwarf.

\begin{figure}
\includegraphics[scale=0.4, angle=0]{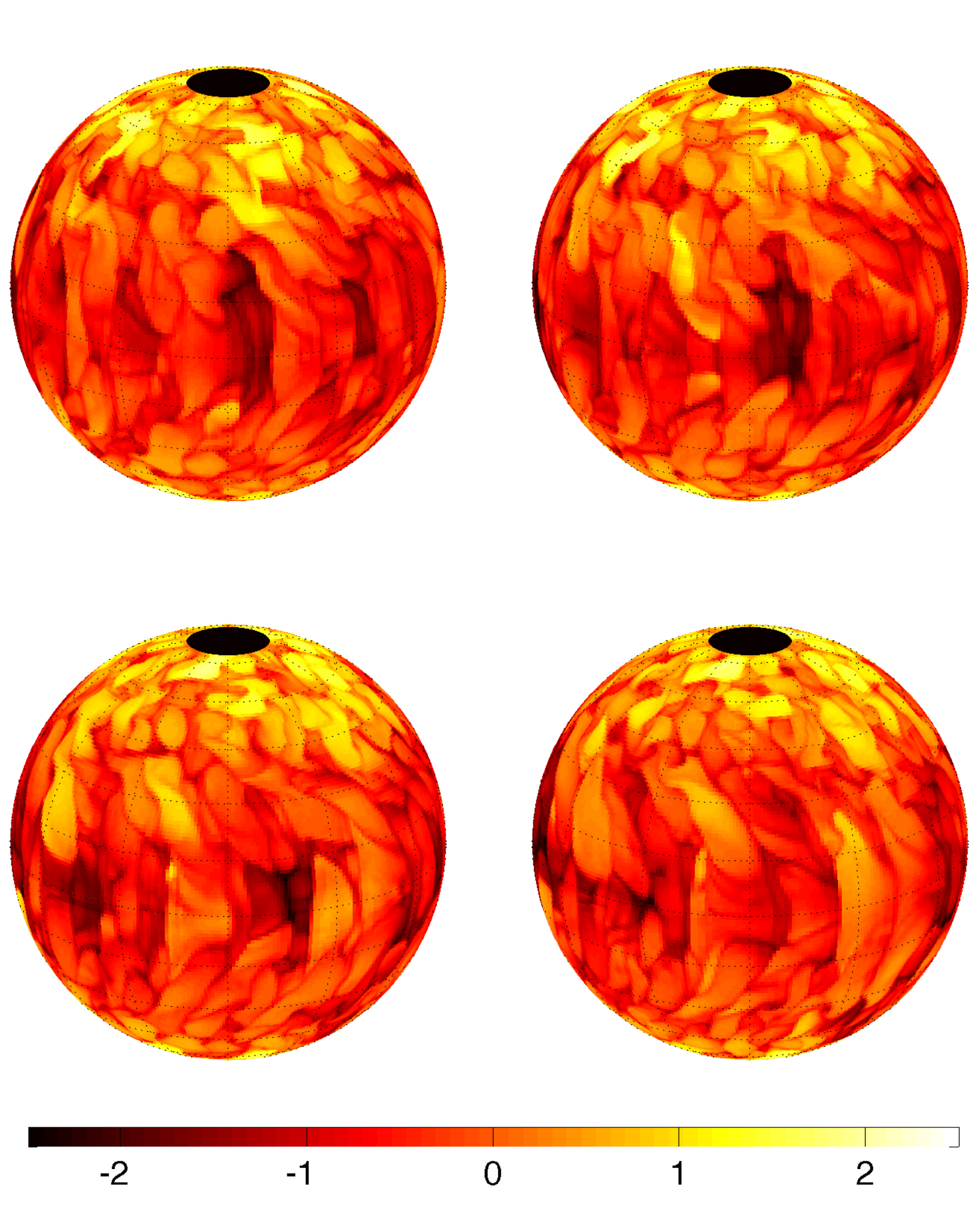}
\caption{Snapshots at different times of the temperature perturbations 
at 1 bar in a single model with rotation period of 10 hours.  Temperature
perturbations are deviations of temperature from the reference state,
in K.  Time separation between frames is 4.8 hours. The full sequence
of model snapshots can be viewed on the authors' websites.}
\label{globe-wind}
\end{figure}

The convective structure exhibits significant temporal variability,
as can be seen in Figure~\ref{globe-wind}.  The figure shows the 
1-bar temperature structure (at the top of the convection zone)
at 4.8-hour intervals in a brown dwarf model with a rotation period
of 10 hours.  Such convective variability should cause significant
variability in the overlying atmosphere, helping to explain the
variability in lightcurve shapes observed in several L/T dwarfs
\citep{artigau-etal-2009, radigan-etal-2012}.  We return to the
dynamics of the stratified atmosphere in Section~\ref{atmosphere}.

\subsection{Large-scale flow organization in the convection zone}
\label{convective-flow-organization}

Here we address the question of  whether the convection zone can
develop organized, large-scale horizontal winds such as fast east-west (zonal) 
jets, since these might play a role in causing differential zonal motion of cloudy
and cloud-free regions in the overlying atmosphere. 

At pressures $\gtrsim 1\,$Mbar, hydrogen metallizes and
magnetohydrodynamic (MHD) effects become important
\citep[e.g.,][]{weir-etal-1996, nellis-etal-1995, nellis-etal-1996,
  nellis-2000, nellis-2006}.  Theoretical arguments and numerical
simulations of convection in electrically conducting spherical
shells---as applied to Jupiter, Earth's outer core, and related
systems---suggest that the Lorentz force acts to brake the large-scale
east-west (zonal) winds when the electrical conductivity is high,
inhibiting jet formation in the metallic region
\citep[e.g.,][]{kirk-stevenson-1987, grote-etal-2000a, busse-2002,
  liu-etal-2008}.  Numerical simulations of dynamo generation in
convecting, rotating fluids at high electrical conductivity have led
to scaling laws for the magnetic field strength of rapidly rotating
planets and convective stars \citep{christensen-aubert-2006,
  christensen-etal-2009, christensen-2010}.  Application of these
these scaling laws to brown dwarfs predict that brown dwarfs
will exhibit strong magnetic fields
\citep{reiners-christensen-2010}.  These dynamo experiments also lead
to scaling laws for the mean flow velocities in the dynamo-generating
region \citep{christensen-aubert-2006, christensen-2010}.  When
heat fluxes, rotation rates, and densities appropriate to typical
L/T-transition dwarfs are adopted ($F\sim 10^5\rm\,W\,m^{-2}$,
$\Omega\sim10^{-3}$--$10^{-4}\rm\,s^{-1}$, and $\rho\sim 1$--$5\times10^4
\rm\,kg\,m^{-3}$), these scaling laws predict typical fluid velocites
of $\sim$0.1--$0.3\rm\,m\,s^{-1}$---similar to estimates from
Equation~(\ref{w-rotate}) under the same assumptions.  Overall,
these results suggest that the flow speeds are weak in the 
metallic interiors of brown dwarfs.  We for now proceed
under the assumption that the large-scale horizontal winds are weak in
the metallic region, and ask what happens in the overlying molecular
envelope.

The emergence of large-scale, organized horizontal temperature
gradients (cf Figures~\ref{globe-temp} and \ref{T-vs-lat}) implies that
the flow will develop large-scale shear of the zonal wind in the direction
along the rotation axis via the thermal-wind equation (\ref{thermal-wind}).
We here write this in the form
\begin{equation}
2\Omega {\partial u\over\partial \hat z} \approx g k_{\rm jet}
{\delta \rho\over\rho} \approx g k_{\rm jet} \alpha
\delta T
\label{thermal-wind2}
\end{equation}
where $\delta \rho$ and $\delta T$ are the characteristic large-scale horizontal
density and temperature differences (on isobars) which occur over
a horizontal wavenumber $k_{\rm jet}$ (between the equator and pole,
for example).  We envision that these horizontal density and temperature
differences result from large-scale organization of the convective
temperature fluctuations, and we therefore equate $\delta T$ in
Equation~(\ref{thermal-wind2}) to that from
Equation~(\ref{deltaT-rotate}).  Doing so yields a characteristic
variation of the zonal wind along $\hat z$ of 
\begin{equation}
\Delta u \approx {k_{\rm jet}\over 2}\int \left({F g \alpha\over \rho c_p\Omega}
\right)^{1/2}\,d\hat z
\label{thermal-wind3}
\end{equation}
The quantity $\alpha/\rho$ in the integrand of (\ref{thermal-wind3})
varies by orders of magnitude from the atmosphere to the deep interior
and must be accounted for.  In contrast, $F$, $g$, and $c_p$ vary
radially by a factor of two or less across the molecular
envelope \citep[see][]{kaspi-etal-2009}, and to a first approximation---here
seeking simply an order-of-magnitude expression---we can  treat them
as constant.  If we furthermore adopt the ideal-gas equation of
state, which is reasonably accurate in the outermost layers, and
assume that the background thermal profile is an adiabat, we
can integrate Equation~(\ref{thermal-wind3}) analytically to obtain the
characteristic difference in zonal wind (along $\hat z$)
between a deep pressure 
$p_{\rm bot}$ and some low pressure $p$:
\begin{equation}
\Delta u \approx {-R k_{\rm jet}\theta\over (1-2\kappa)
|\sin\phi| p_0^{\kappa}}\left({F R\over c_p g\Omega}\right)^{1/2}
\left[{1\over p^{{1\over 2}-\kappa}} - {1\over p_{\rm bot}^{{1\over 2}-\kappa}}
\right]
\label{thermal-wind4}
\end{equation}
where we have used the fact that the pressure variation along $\hat z$
is overwhelmingly dominated by the hydrostatic contribution.  
In Equation~(\ref{thermal-wind4}), $\theta = T(p_0/p)^{\kappa}$
is the potential temperature of the adiabat (that is, a representation
of the entropy of the adiabat), $p_0$ is a reference
pressure (which we take here to be 1 bar), $R$ is the specific gas
constant, $\kappa = R/c_p$,
and the region under consideration has a characteristic latitude $\phi$.

Adopting values appropriate to a typical brown dwarf 
($R=3700\rm\,J\,kg^{-1}\,K^{-1}$,
$\kappa=2/7$, $\theta = 1000\rm \,K$, $\Omega \approx 3\times10^{-4}\rm\,s^{-1}$,
$\phi\approx 30^{\circ}$,
$F\sim10^5\rm\,W\,m^{-2}$, $g\approx 500\rm\,m\,s^{-2}$, and 
$k_{\rm jet} = 1\times10^{-7}\rm\,m$ corresponding to a wavelength
of approximately one Jupiter radius), the equation can be expressed
\begin{equation}
\Delta u \approx 2 \left[\left({{1\rm\,bar}\over p}\right)^{{1\over 2}
-\kappa} - \left({{1\rm\,bar}\over p_{\rm bot}}\right)^{{1\over 2}
-\kappa}\right]\rm m\,s^{-1}.
\label{thermal-wind5}
\end{equation} 

We are interested in the wind shear between the deep interior and the
top of the convection zone, where the pressure is approximately
$p\sim1\rm\,bar$.  Interestingly, when we consider any deep pressure
$p_{\rm bot}\gg p$, the second term in Equation~(\ref{thermal-wind5})
drops out and the equation becomes independent of $p_{\rm bot}$; this
is because the factor $\alpha/\rho$ becomes extremely small at high
pressure, so that almost all of the contribution to $\Delta u$ comes
from the outermost few scale heights of the convection zone---even if
a very deep layer is being considered.  With the adopted parameters,
we then obtain $\Delta u \sim 2\rm\,m\,s^{-1}$ for the difference in
zonal wind (in the direction of $\hat z$) between any deep level and 1
bar.  The implication is that, given the expected temperature
variations associated with convection, the large-scale wind varies by
at most a few $\rm m\,s^{-1}$ along the direction of the rotation
axis.  If the large-scale, zonal-mean horizontal wind is weak in the
deep interior where MHD effects predominate, then it will also be weak
near the top of the convection zone.  If, rather than adopting a
horizontal length scale $2\pi/k_{\rm jet}$ of a Jupiter radius, we
instead adopt a smaller length scale (e.g., 0.1 Jupiter radius,
appropriate to the regional-scale temperature anomalies seen in
Figure~\ref{globe-temp}), we then conclude that horizontal winds of
tens of $\rm m\,s^{-1}$ are possible at the top of the convection
zone.  Despite the uncertainties, these estimates suggest that the
large-scale zonal and meridional wind speeds in the convection zone
are $ < 10^2\rm\,m\,s^{-1}$ for typical brown dwarfs.

\section{DYNAMICS OF THE STRATIFIED ATMOSPHERE}
\label{atmosphere}

We have so far emphasized the convective interior, but infrared
spectra and light curves emerge from pressures that for a wide range
of effective temperature lie within the stratified atmosphere
overlying the convective region \citep[e.g.,][] {burrows-etal-2006b}.
Understanding this stratified region is therefore crucial for
understanding observations.  In particular, the horizontal
temperature differences, wind speeds, and dominant flow length scales
in this layer will control the variability in IR lightcurves, 
and vertical mixing rates will control cloudiness and
chemical disequilibrium.  Here we outline the expected dynamics of
this stratified layer.

\subsection{Qualitative mechanism of atmospheric circulation}

At first glance, it is not obvious that brown dwarfs should exhibit
significant large-scale circulations in their atmospheres.  Because
they receive no external irradiation, the temperature-pressure profiles
in their stratified atmospheres are determined primarily by absorption
of upwelling IR radiation from below. Since the interior entropy of a
brown dwarf varies little with latitude, one might therefore expect
that the radiative-equilibrium temperature profile of the stratified
atmosphere should vary little with latitude, and that---at least at
large scales---the stratified regions will be relatively quiescent.
This constrasts significantly from the tropospheres of most solar
system planets---and hot Jupiters---where differential stellar heating
between equator and pole (or day and night) leads to a thermally
driven atmospheric circulation.

However, the interaction of convective turbulence with the stable
layer on brown dwarfs will perturb the stratified layer and generate a
wide spectrum of atmospheric waves, including gravity waves
\citep[e.g.,][]{goldreich-kumar-1990, freytag-etal-2010} and Rossby
waves.  In solar-system atmospheres, including that of Earth, Jupiter,
Saturn, Uranus, and Neptune, such waves generated in the troposphere
by convection and various instabilities propagate upward into the
stratosphere.  The interaction of these waves with the mean flow---in
particular, the generation, absorption, breaking, and dissipation of
these waves---induces a large-scale circulation in the stratosphere.
Indeed, despite the existence of equator-to-pole radiative (thermal)
forcing in irradiated atmospheres, this {\it mechanical}, wave-induced
forcing is perhaps the dominant driver of the stratospheric
circulation on the Earth and the giant planets \citep[for reviews,
  see, e.g.,][]{andrews-etal-1987, shepherd-2000, shepherd-2003,
  haynes-2005}.  In a similar way, we envision that the breaking,
absorption, and dissipation of convectively generated waves will drive
a large-scale circulation in the stratified atmospheres of brown
dwarfs.

\begin{figure}
\includegraphics[scale=0.38, angle=0]{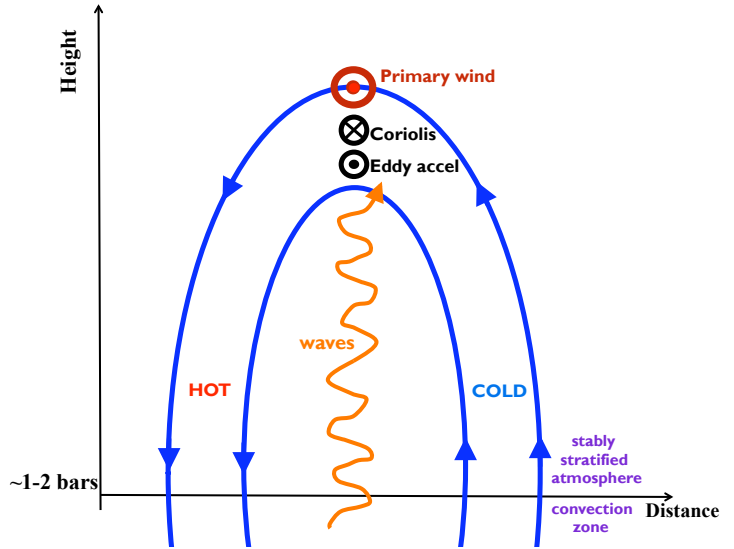}
\caption{Schematic illustration of a wave-driven atmospheric
  circulation, as occurs in the stratospheres of Earth, Jupiter,
  Saturn, Uranus and Neptune and as we propose occurs at
  regional-to-global scales in the stably stratified atmospheres of
  brown dwarfs.  Gravity and Rossby waves (orange) propagate from the
  convective zone into the atmosphere, where they break or dissipate
  and induce an acceleration of the mean wind, here illustrated as a
  vector coming out of the page (black $\odot$ symbol labeled ``eddy
  accel'').  This drives a horizontal wind here also represented as a
  vector coming out of the page (red $\odot$ symbol labeled ``primary
  wind'').  The resulting deviation from geostrophy drives a weak
  secondary circulation (blue contours) in the plane perpendicular to
  the primary wind.  In steady state, the eddy acceleration is
  balanced by a Coriolis acceleration ($\otimes$ symbol, representing
  a vector pointing into the page) associated with the secondary
  circulation.  Vertical motion associated with the secondary circulation
advects entropy vertically, leading to horizontal temperature contrasts
(labelled ``hot'' and ``cold'').  These temperature contrasts are
 precisely those needed to maintain thermal-wind balance with the
 vertical shear of the primary wind.  Scales are uncertain but are
 plausibly thousands to tens of thousands of km horizontally and several
  scale heights vertically.}
\label{wave-driven-circ}
\end{figure}

A variety of nonlinear interactions and feedbacks enhance the ability
of such wave/mean-flow interactions to drive an atmospheric
circulation.  For example, vertically propagating waves are
preferentially absorbed near critical layers where the background flow
speed matches the wave speed; such absorption causes an acceleration
of the mean flow that is spatially coherent.  In Earth's atmosphere,
this effect allows convectively generated waves propagating upward
from the troposphere to drive zonal jets in the stratosphere, a
phenomenon known as the ``Quasi-Biennial Oscillation'' or QBO
\citep{baldwin-etal-2001}.  A similar phenomenon has been observed on
Jupiter \citep{friedson-1999} and has been suggested to occur in hot
stars \citep[e.g.,][]{rogers-etal-2012}.  Likewise, the mixing induced
by breaking Rossby waves is spatially inhomogeneous and naturally
leads to the formation of jets and vortices
\citep[e.g.,][]{dritschel-mcintyre-2008}.  Idealized numerical
experiments of two-dimensional and stratified, three-dimensional,
rapidly rotating flows demonstrate that random turbulent
forcing can generically lead to the generation of large-scale vortices
and jets \citep[e.g.,][]{nozawa-yoden-1997a, huang-robinson-1998,
  marcus-etal-2000, smith-vallis-2001, smith-2004, scott-polvani-2007,
  showman-2007, dritschel-mcintyre-2008, dritschel-scott-2011}.

Regardless of the details of this forcing, the rapid rotation will
dominate the physical structure of such a circulation at large scales
(defined here as say $\gtrsim 10^3\rm\,km$).  The small Rossby numbers
expected at large scales imply that this circulation will be
geostrophically balanced, that is, pressure-gradient forces will
approximately balance Coriolis forces in the horizontal momentum
equation.  Here, we adopt the primitive equations, which are the
standard equations governing atmospheric flows in stably stratified
atmospheres when the horizontal dimensions greatly exceed the vertical
dimensions \citep[for reviews, see][]{pedlosky-1987, vallis-2006,
  showman-etal-2010}.  Using log-pressure as a vertical coordinate,
geostrophy reads
\begin{equation}
fu = -\left({\partial\Phi\over\partial y}\right)_p \qquad 
fv=\left({\partial \Phi\over\partial x}\right)_p
\end{equation}
where $u$ and $v$ are the east-west (zonal) and north-south
(meridional) wind on isobars, $\Phi$ is the gravitational potential on
isobars, $x$ and $y$ are eastward and northward distance,
respectively, and the derivatives are taken on constant-pressure
surfaces.  Here, $f\equiv 2\Omega\sin\phi$ is the Coriolis parameter.
When combined with local hydrostatic balance---valid in the stratified
atmosphere at horizontal scales greatly exceeding vertical
ones---geostrophy implies that the vertical shears of the horizontal
wind relate to the horizontal temperature gradients via the
thermal-wind equation \citep[e.g.,][]{pedlosky-1987, vallis-2006}
\begin{equation}
f{\partial u\over\partial \z} = -R{\partial T\over\partial y}\qquad\qquad
f{\partial v\over\partial \z} = R{\partial T\over\partial x}.
\end{equation}
where $\z \equiv -\ln p$ is the vertical coordinate.
At large scales, then, the development of fast winds in the atmosphere---particularly if the large-scale winds in the convection zone are weak---depends on
the ability of the atmosphere to maintain
horizontal temperature gradients.

What is the nature of this wave-driven circulation?
Figure~\ref{wave-driven-circ} provides a schematic illustration of the
dynamics.  Generally, the acceleration of large-scale horizontal winds
by waves or turbulence induces deviations from geostrophy, leading to
a mismatch between the pressure-gradient and Coriolis forces in the
direction perpendicular to the wind. This unbalanced force drives a
so-called ``secondary circulation'' in the plane perpendicular to the
main geostrophic flow; the Coriolis forces and entropy advection
caused by this circulation act to restore geostrophy.  This standard
mechanism is well understood [see \citet{haynes-etal-1991} for theory,
  and \citet[][pp.~100-107]{james-1994} or
  \citet[][pp.~313-327]{holton-2004} for brief reviews] and provides
the dynamical link between the horizontal winds, temperature
perturbations, and vertical velocities.

\subsection{Quantitative model of atmospheric circulation}
\label{sec: windtheory}

Although the amplitude of the wave driving in brown-dwarf atmospheres
remains unknown, the above dynamical arguments allow us to determine
the relationship between the horizontal winds, temperature contrasts,
and vertical velocities {\it as a function} of the wave-driving
amplitude. We here construct a simple analytic theory of this
atmospheric circulation, treating the wave-driving amplitude as a free
parameter.  The model is approximate and makes a number of simplifying
assumptions in the spirit of exposing the dynamics in the simplest
possible context.  As a result, the model is not expected to be accurate in
quantitative detail.  Rather, the goal is to broadly illustrate the
types of physical processes governing the atmospheric circulation on
brown dwarfs, and to obtain order-of-magnitude estimates for the
horizontal temperature perturbations and wind speeds, quantities
important in shaping the observables.

In steady state,
the momentum balance in the direction parallel to the geostrophic wind
reads, to order of magnitude\footnote{Suppose, for concreteness, that the
  dominant geostrophic flow consists of zonal jets, as exist on
  Jupiter.  The significant zonal symmetry of such jets suggests
  decomposing the flow into zonal-mean and deviation (eddy)
  components, $A= \overline{A} + A'$.  By expanding the zonal momentum
  equation and zonally averaging, we obtain the Eulerian-mean equation
  for the evolution of the zonal-mean flow, $\overline{u}$, over time
  (adopting Cartesian geometry for simplicity)
\begin{equation}
{\partial \overline{u}\over\partial t} = -\overline{v}{\partial 
\overline{u}\over\partial y} - \overline\varpi {\partial \overline{u}
\over\partial \z} + f\overline{v} - {\partial(\overline{u'v'})
\over\partial y} - e^{\z}{\partial (e^{-\z}\overline{u'\varpi'})\over\partial \z}
\label{eulerian-zonal-mean}
\end{equation}
where $\varpi = d\z/dt = -d\ln p/dt$ is the vertical velocity in
log-pressure coordinates.  Thus, the absorption, breaking, or
dissipation of waves can drive a mean flow, $\overline{u}$.  Scaling
analysis of this equation immediately shows that, on the righthand
side, the first and second terms are both order $Ro$ smaller than the
third term.  In steady state, then, the balance in a geostrophic flow
is between the eddy-driven accelerations and the Coriolis force
associated with a mean meridional circulation, i.e.,
\begin{equation}
f\overline{v} \approx  {\partial(\overline{u'v'})
\over\partial y} + {\partial (\overline{u'\varpi'})\over\partial \z}.
\label{eulerian-zonal-mean2}
\end{equation}
If the flow consists predominantly of large vortices
rather than zonal jets, one can alternately adopt a cylindrical coordinate system
centered on a vortex, where $u$ is the azimuthal flow around the
vortex, $v$ is the radial velocity (toward/away from the vortex
center), and the eddy-mean-flow composition denotes an {\it azimuthal}
mean around the vortex (overbars), and deviations therefrom (primes).  
Azimuthally averaging the azimuthal momentum equation then leads to 
relationships analogous to (\ref{eulerian-zonal-mean}) and
(\ref{eulerian-zonal-mean2}).  In either case, the equation
can be expressed, to order of magnitude, as Equation~(\ref{zonal-momentum}).}
\begin{equation}
f\overline{v} \sim {\cal A}
\label{zonal-momentum}
\end{equation}
where $\overline{v}$ is the horizontal flow perpendicular to the main
geostrophic flow and $\cal{A}$ is the characteristic magnitude of the
eddy-induced acceleration of the mean flow, due to breaking,
absorption, or dissipation of gravity or Rossby waves.  What this
equation implies is that the wave interactions with the mean flow
drive a so-called secondary circulation, $\overline{v}$ (meridional in
the case of zonal jets, radially toward or away from the vortex center
in the case of a large vortex).  See Figure~\ref{wave-driven-circ}
for the conceptual picture.

The associated vertical velocity can be obtained from the continuity
equation, which for the primitive equations in log-pressure
coordinates reads
\begin{equation}
{\partial \overline{v}\over\partial y} + e^{\z} {\partial\over
\partial \z}(e^{-\z}\overline{\varpi}) = 0
\end{equation}
which we can approximate to order of magnitude as
\begin{equation}
\overline{v}l \sim {\overline{\varpi} \over \Delta\z},
\label{continuity-pe}
\end{equation}
where $\overline{\varpi}$ is the characteristic vertical velocity (in units of
scale heights per sec), $\Delta \z$ is the vertical scale of the
circulation (in units of scale heights), and $\l$ is the dominant
horizontal wavenumber of the circulation. Equations
(\ref{zonal-momentum}) and (\ref{continuity-pe}) imply that
\begin{equation}
\overline{\varpi} \sim \overline{v} l \Delta\z \sim {l {\cal A}
\Delta\z\over f}.
\label{vert-pe}
\end{equation}
Thus, wave interactions with the mean flow drive large-scale vertical
motions.

These wave-driven, large-scale vertical motions will advect
entropy vertically, leading to the existence of horizontal 
temperature variations on isobars.  These are in fact exactly
the temperature differences needed to maintain the wave-driven 
geostrophic flow in thermal-wind balance.  To quantify, consider 
the thermodynamic energy equation, which can be expressed as
\begin{equation}
{\partial T\over \partial t} + \mathbf{v}_h\cdot\nabla_h T
- \varpi{H^2N^2\over R} = {q\over c_p}
\end{equation}
where $T$ is temperature, $\mathbf{v}_h$ the horizontal velocity,
$\nabla_h$ is the horizontal gradient operator,
$H$ the scale height, $N$ the Brunt-Vaisala frequency, $q$ the specific heating
rate, and $c_p$ the specific heat. In a statistical steady state, we envision
a primary balance between the radiative heating/cooling (righthand side)
and vertical advection (right term on left side).  If isentrope
slopes are sufficiently large, the horizontal mixing may also contribute
via the term $\mathbf{v}\cdot\nabla T$.  We write this balance as
\begin{equation}
-\varpi{H^2N^2\over R} = {q\over c_p} - \mathbf{v}_h\cdot \nabla_h T.
\end{equation}
We parameterize radiative heating/cooling as Newtonian relaxation of
the temperature toward the radiative-equilibrium state, expressed as
$q/c_p = (T_{\rm eq} - T)/\tau_{\rm rad}$, where $T_{\rm eq}(\z)$ is
the radiative-equilibrium temperature profile and $\tau_{\rm rad}$ is
a specified radiative time constant.  Since brown dwarfs receive no
external irradiation, $T_{\rm eq}$ is to zeroth order independent of
latitude and longitude.  To order-of-magnitude, the characteristic deviation of
temperature from its local radiative equilibrium, $T(\z)-T_{\rm
  eq}(\z)$, is comparable to the characteristic horizontal temperature
difference on isobars, $\Delta T_{\rm horiz}$.  We also parameterize
the meridional eddy mixing as a diffusive process, with eddy
diffusivity $D$.  To order-of-magnitude, we thus have
\begin{equation}
\overline{\varpi}{H^2N^2\over R} = {\Delta T_{\rm horiz} \over \tau_{\rm rad}} 
+ D l^2\Delta T_{\rm horiz}.
\label{energy}
\end{equation}
The physical interpretation is that vertical advection (left
side) attempts to {\it increase} the horizontal temperature contrasts,
whereas radiation and meridional eddy mixing (right side) both attempt
to {\it decrease} the horizontal temperature contrasts.  Here,
$\overline{\varpi}$ and $\Delta T_{\rm horiz}$ refer to characteristic
magnitudes and are defined positive.  Importantly, the two terms on
the right side have the same sign, since they both act in the same
direction, namely, to damp temperature differences.\footnote{Breaking
gravity waves will cause a vertical mixing that might be represented
as a {\it vertical} diffusion of entropy, leading to an additional
source term in Equation~(\ref{energy}).  Only horizontal {\it variations}
in the amplitude of this mixing will act to alter $\Delta T_{\rm horiz}$.  
A priori, it is not clear how such variations will correlate
with the overturning circulation nor how to parameterize them in
the context of Equation~(\ref{energy}).  Since our goal is to
describe the dynamics of the wave-driven circulation in the simplest
possible context, we therefore forgo any inclusion of this vertical mixing term
here, with the understanding that more realistic models of the
large-scale circulation
will probably have to account for it.}

Substituting Eq.~(\ref{vert-pe}) into Eq.~(\ref{energy}) yields
\begin{equation}
{l {\cal A} \Delta\z H^2N^2\over fR}\sim
\Delta T_{\rm horiz}\left({1\over \tau_{\rm rad}} + Dl^2\right)
\end{equation}
which can readily be solved to yield an expression for 
the meridional temperature difference in terms of ``known'' parameters:
\begin{equation}
\Delta T_{\rm horiz} \sim {l {\cal A} \Delta\z H^2N^2
\over f R \left({1\over \tau_{\rm rad}} + Dl^2\right)}.
\label{temp-diff}
\end{equation}

We have yet to use the meridional momentum balance (or radial
momentum balance in the case of a vortex), and doing so
will allow us to solve for the zonal wind several scale heights
above the radiative-convective boundary.  To order-of-magnitude,
the thermal-wind equation implies
\begin{equation}
\Delta u \sim {R l \Delta T_{\rm horiz}\Delta\z\over f}
\label{thermal-wind-pe}
\end{equation}
where $\Delta u$ is the characteristic difference between the 
wind speed at the radiative-convective boundary and some level of
interest, say at the mean IR photosphere. If the former is small as
suggested in Section~\ref{convective-flow-organization}, $\Delta u$
would approximately give the actual wind speed at levels above the
radiative-convective boundary.  Inserting
Eq.~(\ref{temp-diff}) into Eq.~(\ref{thermal-wind-pe}), we obtain
\begin{equation}
\Delta u \sim {l^2 {\cal A} \Delta\z^2 H^2N^2
\over f^2 \left({1\over \tau_{\rm rad}} + Dl^2\right)}.
\label{u-diff}
\end{equation}

Together, Equations~(\ref{vert-pe}), (\ref{temp-diff}), and (\ref{u-diff})
provide the expressions we seek
for the vertical velocities, horizontal temperature differences,
and horizontal wind speeds as a function of ${\cal A}$, $l$, and
parameters that are either known or can be estimated.   The 
Coriolis parameter, $f$, follows directly from the rotation period.
For a brown dwarf of a given effective temperature and gravity,
1D radiative-transfer models allow estimates of the vertical
temperature-pressure profile \citep[e.g.,][]{marley-etal-1996,
 marley-etal-2002, marley-etal-2010, burrows-etal-1997, burrows-etal-2006}
and hence $HN$.  Since
the infrared photosphere is typically 1--3 scale heights above
the radiative-convective boundary \citep{burrows-etal-2006}, values
of $\Delta \z \sim 1$--3 are most appropriate.

What sets the dominant horizontal length scale of the flow,
represented in the above theory by the wavenumber $l$?  One
possibility is the Rhines scale, given by $({\Delta u/\beta})^{1/2}$
where $\beta$ is the derivative of the Coriolis parameter with
northward distance $y$.  This is generally the scale at which Rossby
waves impose anisotropy on the flow, and in many systems, it is the
energy-containing scale.  The Rhines scale controls the jet widths on
Jupiter, Saturn, Uranus, and Neptune
\citep[e.g.,][]{cho-polvani-1996a} as well as in a wide range of
numerical simulations of stratified, rotating turbulence \citep[for a
  review, see][]{vasavada-showman-2005}.  For typical brown dwarf
rotation rates and the wind speeds estimated in
Figure~\ref{windtheory}, this yields $l\sim
3$--$6\times10^{-7}\rm\,m^{-1}$, corresponding to horizontal
wavelengths of 10,000 to 20,000 km.  Another possibility is that $l$
results from an interaction of turbulent energy transfers with the
strong radiative and/or frictional damping.  Stratified flows forced
at small scales tend to exhibit upscale energy cascades, and a
competition between the rate of upscale energy transfer and the
radiative damping timescale then determines the dominant length scale.
This possibility is at present difficult to quantify. Given the
uncertainties, we simply adopt plausible values for $l$ here, and
leave a detailed investigation for future work.

It is also worthwhile expressing our solutions in terms of the power
exerted by the waves in driving the large-scale circulation.  The
characteristic power per mass exerted by the waves in driving the
large-scale circulation is approximately ${\cal A}\Delta u$, and the
power per unit horizontal area is ${\cal A}p \Delta u/g$, where $p$ is
the pressure at the radiative-convective boundary.  Defining a
dimensionless efficiency $\eta$, corresponding to the fraction of the
brown-dwarf heat flux that is used to drive the atmospheric
circulation, we then have $\eta F \sim {\cal A}p\Delta u/g$, where $F$
is the heat flux radiated by the brown dwarf.  Using this constraint,
the solutions become
\begin{equation}
\Delta T_{\rm horiz} \sim  \left({\eta g F \over p}\right)^{1/2}   {H N\over R\left(
{1\over \tau_{\rm rad}} + Dl^2\right)^{1/2}},
\label{temp-eta}
\end{equation}
\begin{equation}
\Delta u \sim \left({\eta g F \over p}\right)^{1/2} {l \Delta\z H N\over f
\left({1\over \tau_{\rm rad}} + Dl^2\right)^{1/2}},
\label{wind-eta}
\end{equation}
and
\begin{equation}
\varpi \sim \left({\eta g F \over p}\right)^{1/2} {1\over
  HN}\left({1\over \tau_{\rm rad}} + Dl^2\right)^{1/2}.
\label{vert-eta}
\end{equation}
If we adopt a blackbody flux, $F=\sigma T^4$, where $T$ is the typical
photospheric temperature and $\sigma$ is the Stefan-Boltzmann constant, 
and express the radiative time constant
as \citep{showman-guillot-2002}
\begin{equation}
\tau_{\rm rad} \sim {p c_p\over 4 g\sigma T^3}
\label{tau-rad}
\end{equation}
then Equations~(\ref{temp-eta}), (\ref{wind-eta}), and (\ref{vert-eta}) can
be expressed as functions of temperature and dimensionless wave-driving efficiency
(here for simplicity neglecting the horizontal diffusion term),
\begin{equation}
\Delta T_{\rm horiz} \sim \left({\eta c_p T\over 4}\right)^{1/2} {NH\over R},
\label{temp-eta-temp}
\end{equation}
\begin{equation}
\Delta u \sim \left({\eta T c_p\over 4}\right)^{1/2} {l\Delta \z HN\over f},
\label{u-eta-temp}
\end{equation}
and
\begin{equation}
\varpi \sim {2\eta^{1/2} g T^{7/2}\sigma\over c_p^{1/2} p HN}\sim {2 (\eta T c_p)^{1/2}
\over HN\tau_{\rm rad}}.
\label{vert-eta-temp}
\end{equation}

Noting that $HN$ is the approximate horizontal phase speed of
long-vertical-wavelength gravity waves, inspection of
(\ref{temp-eta-temp})--(\ref{vert-eta-temp}) makes clear that, to
within factors of order unity,
\begin{itemize}
\item $\Delta T_{\rm horiz}/T$ is $\eta^{1/2}$ times the ratio of the
  gravity wave speed to the sound speed,
\item $\Delta u$ over the sound speed is $\eta^{1/2}$ times the ratio
  of the Rossby deformation radius, $L_D = \Delta\z HN/f$, to the
  dominant horizontal length scale of the flow, and
\item $\varpi$ is $\eta^{1/2}\tau_{\rm rad}^{-1}$ times the ratio of
  the sound speed to the gravity wave speed (in other words, the
time for the flow to advect vertically over a scale height is
$\eta^{-1/2}\tau_{\rm rad}$ times the ratio of the gravity wave speed
to the sound speed).
\end{itemize}
For an isothermal, ideal-gas atmosphere, $HN = R\sqrt{T/c_p}$, and the
ratio of $HN$ to the sound speed is $\sqrt{\kappa(1-\kappa)}$, which
is $\sim$0.4 for an H$_2$ atmosphere with $\kappa = 2/7$.  Thus,
the ratio of the gravity wave speed to the sound speed is order unity.
These arguments imply that, for small wave-driving efficiencies
($\eta \ll 1$), the fractional horizontal temperature differences will be
small, the horizontal wind speeds will be much less than the sound speed, and
the time for air to advect vertically over a scale height will be
much longer than the radiative time constant.

\subsection{Application of the theory to giant planets and brown
dwarfs}

Detailed numerical simulations of convection impinging on a stable
layer will be necessary to quantify the value of $\eta$, but several
previous studies provide constraints.  In a theoretical investigation
of convection interacting with an overlying isothermal radiative zone,
\citet{goldreich-kumar-1990} found that the fraction of the convective
heat flux converted into gravity waves is approximately the Mach
number associated with the convection, which may be $\sim$0.01 for
typical brown dwarfs.  This presumably provides an upper limit on
$\eta$ since only a fraction of the energy converted to waves is
actually used to drive a large-scale circulation.  Rough estimates for
the Earth's stratosphere\footnote{\label{earth-jup-efficiency} In
  Earth's mid and high latitudes, upwardly propagating waves lead to
  typical accelerations of the zonal-mean zonal wind of order ${\cal
    A}\sim 10^{-5}\rm\,m\,s^{-2}$ in the stratosphere \citep[][Chapter
    13] {andrews-etal-1987, vallis-2006}.  Typical zonal wind speeds
  in these regions are $\Delta u\sim 20\rm\,m\,s^{-1}$.  Adopting a
  pressure $p\sim 0.1\rm\,bar$ representative of the lower
  stratosphere, this implies a power per area driving stratospheric
  motions of ${\cal A}p\Delta u/g \sim 0.2\rm\,W\,m^{-2}$.  Since
  Earth's global-mean radiated flux is $F=240\rm\,W\,m^{-2}$, the
  implied efficiency is $\eta \sim {\cal A}p\Delta u/(F g) \sim
  10^{-3}$.  For Jupiter, observational diagnosis of stratospheric
  heating patterns imply typical eddy accelerations of the zonal-mean
  zonal wind exceeding ${\cal A}\sim 10^{-6}\rm m\,s^{-2}$ in the
  lower stratosphere \citep{west-etal-1992, moreno-sedano-1997}.
  Given a typical stratospheric wind speed of $\sim$$20\rm \,m\,s^{-1}$,
  lower stratospheric pressure of 0.1 bar, and a radiated flux of
  $F\sim 14\rm\,W\,m^{-2}$, this again implies an efficiency
  $\eta\sim10^{-3}$.}  indicate that waves drive a circulation with an
efficiency $\eta\sim10^{-3}$.  Information is limited for Jupiter but
likewise suggests $\eta\sim 10^{-3}$
(footnote~\ref{earth-jup-efficiency}).  While future work is clearly
needed, these estimates suggest that values of $\eta$ ranging from
$10^{-4}$ to $10^{-2}$ may be appropriate to brown dwarfs.

We first test the theory on Jupiter's stratospheric circulation.
Jupiter's mean stratospheric temperature profile rises from the
$\sim$110-K tropopause minimum at 150 mbar to $\sim$$170\rm\,K$ at 1
mbar pressure.  Voyager and Cassini observations show that, throughout
this pressure range, the temperature and zonal wind vary on
characteristic horizontal (meridional) scales of $\sim$$10^4\rm\,km$.
On these scales, temperatures vary by $\sim$3--$5\rm\,K$ at most
latitudes, reaching $10\rm\,K$ at a few latitudes and pressures
\citep{simon-miller-etal-2006}.  Analysis of these observations
indicates that, from 1--100 mbar, zonal winds are
$\sim$20--$30\rm\,m\,s^{-1}$ over most of the planet but reach
$\sim$$130\rm\,m\,s^{-1}$ in specific latitude strips including the
equator and a narrow jet at $23^{\circ}$N
\citep{simon-miller-etal-2006}.  Vertical velocities are less certain
but have been estimated at $\sim$$10^{-5}\rm\, m\,s^{-1}$ throughout
much of the stratosphere, reaching speeds of
$\sim$$3\times10^{-4}\rm\,m\,s^{-1}$ at high latitudes
\citep{moreno-sedano-1997}.

To apply Equations~(\ref{temp-eta-temp}), (\ref{u-eta-temp}), and
(\ref{vert-eta-temp}) to Jupiter, we adopt
$f=2.4\times10^{-4}\rm\,s^{-1}$ (appropriate to $45^{\circ}$
latitude), $R=3700\rm\,J\,kg^{-1}\,K^{-1}$, $p=0.1\rm\,$bar,
$c_p=1.3\times10^4\rm\,J\,kg^{-1}\,K^{-1}$, and evaluate $NH$ using an
isothermal background temperature profile with a temperature of
$165\rm\,K$.  Using a length scale of $10^4\rm\,km$ (implying
$l=6\times10^{-7}\rm\,m^{-1}$) and an efficiency $\eta\sim 10^{-3}$
(see footnote \ref{earth-jup-efficiency}), our theory predicts $\Delta
T_{\rm horiz}\sim 3\rm\,K$, $\Delta u\sim 50\rm\,m\,s^{-1}$, and
$\overline{\varpi} \sim 1\times10^{-8}\rm\,s^{-1}$, which for a scale
height of $20\rm\,km$ implies a vertical velocity of
$\overline{\varpi}H \sim 2\times10^{-4}\rm\,m\,s^{-1}$.  The predicted
meridional temperature contrasts and horizontal wind speeds match the
observations reasonably well.  The predicted vertical velocity lies
close to the upper end of the observationally inferred range,
suggesting that our theoretical estimate may be several times larger
than the actual global-mean vertical velocity (a mismatch that may
result from the crudity of our parameterization of radiative
heating/cooling).  Over and above the specific numerical comparisons,
it is worth emphasizing that observational analysis of Jupiter's
stratospheric circulations supports the overall dynamical framework
adopted here; the primary complication is that Jupiter (unlike brown
dwarfs) exhibits a latitudinal gradient of stellar irradiation, and
thus its stratosphere exhibits aspects of both thermal and
mechanically driven circulations.  Despite this complication, the
comparison is encouraging, and gives us confidence in applying the
theory to brown dwarfs and directly imaged EGPs.

\begin{figure}
\includegraphics[scale=0.65, angle=0]{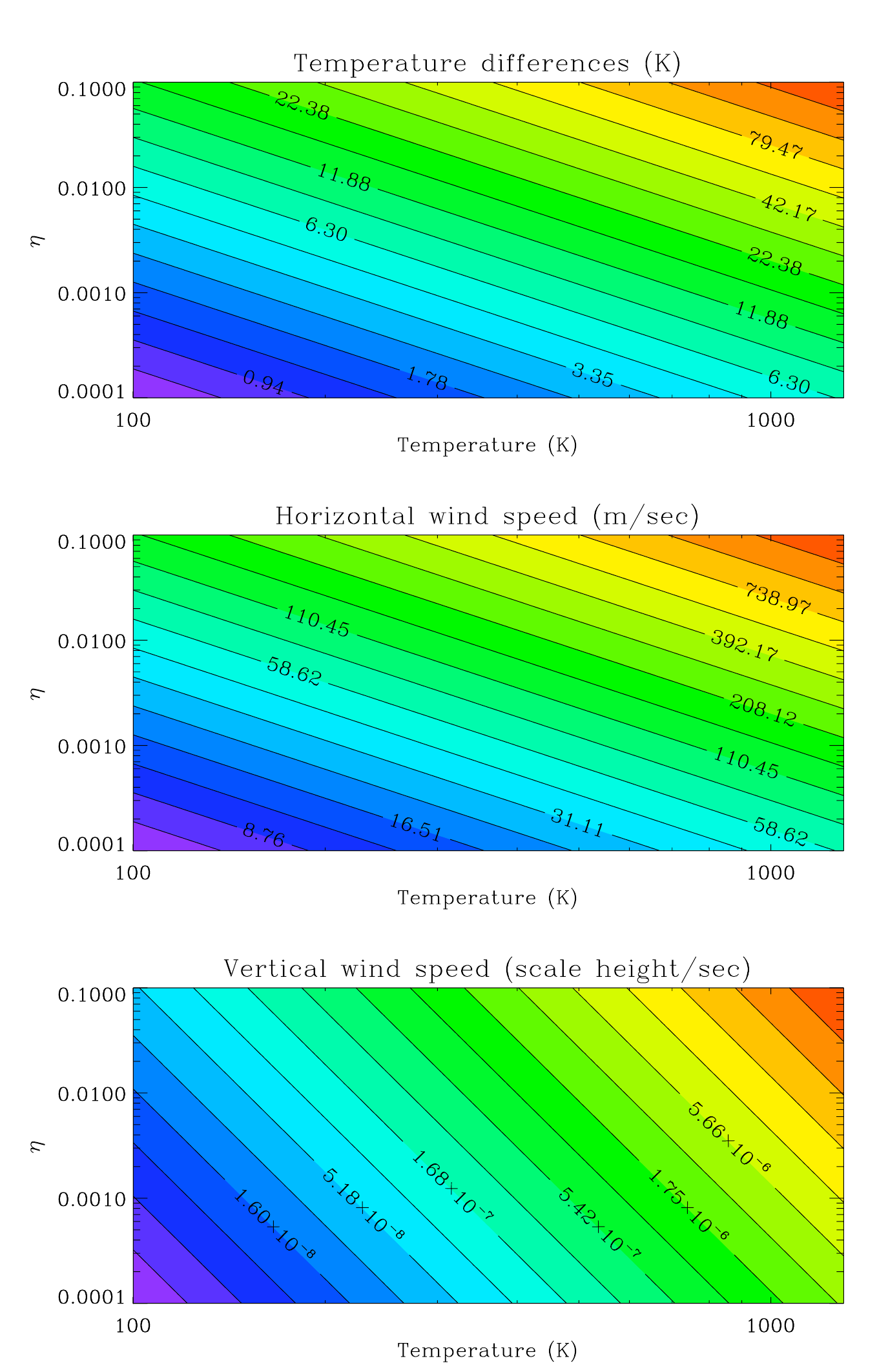}
\caption{Characteristic horizontal temperature differences (top),
  horizontal wind speeds (middle), and vertical velocities (bottom)
  from our solutions for large-scale, wave-driven circulation in the
  stratified atmosphere. Plotted as a function of temperature and
  dimensionless efficiency by which the waves drive the atmospheric
  circulation.  Plotted values adopt a horizontal wavenumber
  $l=6\times10^{-7}\rm\,m^{-1}$ (corresponding to a horizontal
  wavelength of $10^4\rm\,km$) and an isothermal vertical background
  temperature profile (for which $NH = R\sqrt{T/c_p}$).  Other
  parameter values are $R=3700\rm\,J\,kg^{-1}\,K^{-1}$,
  $\Delta\z = 2$ (implying a circulation two
  scale heights tall), and $f=5\times 10^{-4}\rm\,s^{-1}$ (corresponding to a
  rotation period of 5 hours).}
\label{windtheory}
\end{figure}

We next apply the theory to brown dwarfs.  Figure~\ref{windtheory}
shows the solutions (\ref{temp-eta-temp}), (\ref{u-eta-temp}), and
(\ref{vert-eta-temp}) as a function of temperature and $\eta$ for
parameter values representative of brown dwarfs.  The dominant
horizontal length scale of the flow is poorly constrained, and we here
adopt a wavelength of $10^4\rm\,km$, similar to the jet widths on
Jupiter and Saturn.  For simplicity, we parameterize the mean
atmospheric temperature profile in the radiative zone as vertically
isothermal, which implies that $NH = R\sqrt{T/c_p}$.  The magnitude of
the horizontal diffusivity $D$ is likewise poorly constrained;
nevertheless, the baroclinic eddy fluxes that $D$ parameterizes should
be small if isentropes are nearly flat, which will be the case in
strongly stratified regions if
horizontal temperature perturbations are small compared to the mean
temperature.  We here set $D=0$, yielding an upper limit on the
amplitude of the horizontal temperature and winds.

Key points are as follows.  First, at the $\sim$$1000\rm\,K$
temperatures of typical L/T dwarfs, the predicted horizontal
temperature perturbations on isobars are $\sim$5--$50\rm\,K$ for
plausible efficiencies.  Notably, in the $D\to 0$ limit, the
temperature perturbations at a given $\eta$ are independent of
lengthscale and rotation rate.  To see why, note that the steady state
results from a balance between forcing (energy input) and damping
(energy removal).  Energy input is represented by $\eta$; in the model
assumed here, energy damping results solely from loss of potential
energy by radiative relaxation, and for a given value of $\tau_{\rm
  rad}$, this energy damping depends on $\Delta T_{\rm horiz}$ but is
independent of the rotation rate and flow lengthscales.  The
temperature perturbations in Figure~\ref{windtheory} are much less
than the mean temperature, which, given plausible stratifications,
suggests that isentropes are relatively flat.  This provides tentative
{\it a posteriori} justification for neglecting $D$, as horizontal
energy fluxes due to baroclinic eddies---which is what $D$
represents---tend to be modest under such conditions.

Second, Figure~\ref{windtheory} suggests that large-scale, organized
horizontal wind speeds of tens to hundreds of $\rm m\,s^{-1}$ are
plausible for L/T dwarfs.  The range of speeds predicted here are
similar to those suggested from analysis of brown-dwarf variability
\citep{artigau-etal-2009, radigan-etal-2012}, which range from tens to
hundreds of $\rm \,m\,s^{-1}$.  Unlike temperature differences, the
predicted wind speeds depend on length scale and rotation rate.  This
is most clearly seen from the thermal-wind relationship
(\ref{thermal-wind-pe}): for a given $\Delta T_{\rm horiz}$, shorter
length scales imply larger horizontal pressure-gradient forces,
requiring larger wind speed to maintain geostrophy.  Likewise, a
slower rotation rate implies weaker Coriolis forces for a given wind
speed; to balance a given pressure-gradient force, faster winds would
be needed at slower rotation rates.  At large scales, these winds will
comprise stratified, geostrophically balanced turbulence, possibly
organized into coherent structures such as vortices or zonal jets.

Third, for plausible efficiencies, our theory predicts large-scale
vertical velocities of $\sim$$10^{-6}$ to $10^{-5}$ scale heights per
second (Figure~\ref{windtheory}).  Typical times for air parcels to
ascend or descend over a scale height are therefore
$\sim$$10^5$--$10^6\rm\,s$.  For a typical brown dwarf scale height of
$7\rm\,km$, this implies vertical velocities of
$\sim$0.01--$0.1\rm\,m\,s^{-1}$.  Importantly, this motion is not
diffusive but is coherent vertically, comprising the
ascending/descending branches of a large-scale overturning
circulation.  We emphasize that these vertical motions occur
within the {\it stratified} regions above the radiative-convective
boundary and are distinct from the convection below the
radiative-convective boundary.

Such a wave-driven overturning circulation differs fundamentally from
a thermally driven circulation such as a Hadley cell.  Because the
horizontal temperature differences result primarily from vertical
advection of the background vertical entropy gradient, the model
implies that, on isobars, the ascending regions are {\it cold} and
descending regions are {\it hot}.  This thermally indirect
configuration results from the fact that the circulation is
mechanically rather than thermally driven.  Such thermally indirect,
wave-driven circulations are generic features of planetary upper
tropospheres and stratospheres.  On Earth, the main overturning
circulation in the stratosphere, the Brewer-Dobson circulation, is
just such a phenomenon \citep[for reviews, see][Chapter 13]
{haynes-2005, vallis-2006}.  Likewise, in Jupiter's upper troposphere,
observations indicate the existence of ascending motion in the cold,
cloudy ``zones'' and descending motion in the warmer, less-cloudy
``belts'' \citep{gierasch-etal-1986, west-etal-1992}.  Of course, in
the brown-dwarf context, we have neglected the potentially important
roles of latent heating and radiative feedbacks involving clouds,
which may complicate the picture.

In summary, our theory predicts that the stratified atmospheres of
brown dwarfs and giant planets will generically contain geostrophic
flows accompanied by a large-scale overturning circulation, both
driven by the absorption of waves propagating from near the
radiative-convective boundary.  This overturning circulation comprises
coherent, large-scale horizontal {\it and vertical} motions despite
the stratified thermal structure.  The vertical entropy advection
caused by this overturning motion generates large-scale horizontal
temperature differences.

\section{Observational implications}
\label{observables}

\subsection{Clouds and lightcurve variability}

Our results suggest horizontal temperature variations (on isobars) in
the stratified atmosphere up to $\sim$$50\rm\,K$, and even in the
absence of clouds, these variations could cause lightcurve
variability.  The blackbody flux from a region with a photospheric
temperature $T+\Delta T_{\rm horiz}$ differs from that in a region of
photospheric temperature $T$ by $\Delta F\sim 4\sigma T^3\Delta T_{\rm
  horiz}$, implying that the fractional broadband flux variations
emitted by different regions of the brown dwarf, due to thermal
variations alone, are $\Delta F/F \sim 4\Delta T_{\rm horiz}/T$.  For
a mean temperature $T\sim1000\rm\,K$, the fractional temperature
variations are $\sim$0.005-0.05, and $\Delta F/F \sim 0.02$--0.2.  Of
course, if the atmospheric features are much smaller than the radius
of the brown dwarf, these variations will largely cancel and produce
minimal variation in the disk-integrated flux on rotational
timescales.  But if the dominant length scales approach a brown-dwarf
radius, these results suggest that thermal variations alone could
produce variations in the flux of up to a few percent on rotation
timescales.  This might be relevant to brown dwarfs where cloud
condensation is not expected to occur at and above the photosphere
(e.g., on Y and late T dwarfs).

Our models suggest that, when cloud condensation levels lie in the
atmosphere, patchy clouds will form that could lead to significant
variability.  \citet{freytag-etal-2010} predicted cloud patchiness at
very small scales due to fluctuating gravity waves, but such
small-scale fluctuations will cancel out in a disk average and thus
will not produce lightcurve variations.  In contrast, our models
predict {\it large-scale}, vertically coherent regions of ascent and
descent associated with geostrophically balanced, turbulent structures
such as large vortices, with time-mean vertical velocities up to
$\sim$$0.1\rm\,m\,s^{-1}$.  Importantly, these vertical motions occur
despite the fact that the atmosphere is stably stratified.  Regions of
ascent promote cloud formation, whereas regions of descent transport
dry air from above and inhibit cloud formation---just as occurs with
the cloudy zones and less-cloudy belts on Jupiter.  Near typical
photospheric pressures of $\sim$1 bar, the continuum particle-settling
regime is approximately valid; if particles settle following Stokes
flow, then the timescale to settle vertically by a scale height
exceeds $\sim$$10^5$--$10^6\rm\,s$ for particle sizes less than about
$1\,\mu$m.\footnote{For temperatures and molecular viscosities
  appropriate to L/T dwarfs, gas kinetic effects cause a significant
  deviation from Stokes fall speeds only at pressures less than
  $\sim$$0.1\rm\,bar$, indicating that Stokes settling represents a
  reasonable approximation near the IR photosphere. The Stokes
  settling velocity is $w_{\rm settle} = 2\Delta\rho g a^2/9\mu$,
  where $\Delta\rho$ is the density difference between the particles
  and air, $g$ is gravity, $a$ is particle radius, and $\mu$ is the
  dynamic viscosity of H$_2$ air.  Equating this fall speed to the
  mean vertical wind speed $H\overline{\varpi}$ leads to the condition
  that particles can remain suspended if the particle size
\begin{equation}
a\lesssim \left({9\mu H\overline{\varpi}\over2\Delta\rho g}\right)^{1/2}.
\end{equation} 
Adopting a viscosity of $2\times10^{-5}\rm\,Pa\,s$ appropriate to
H$_2$ at 1000 K \citep{ackerman-marley-2001},
$\Delta\rho=3000\rm\,kg\,m^{-3}$, $g=500\rm\,m\,s^{-1}$, $H =
5\rm\,km$, and $\overline{\varpi} = 10^{-6}$--$10^{-5}\rm\,s^{-1}$
yields the condition $a\lesssim 0.5$--$2\,\mu$m for particles to
remain suspended.}  Thus, given our predicted vertical velocities,
particles with radii less than $\sim$$1\,\mu$m can remain suspended in
regions of large-scale ascent, with regions of large-scale descent
exhibiting lower cloud abundances.  Such large-scale cloud patchiness
can lead to significant lightcurve variability as cloudy and
relatively cloud-free regions rotate into and out of view.

How does the vertical scale of any cloud layers relate to that
of the circulation itself?   As mentioned previously,
the atmospheric circulation is likely to extend coherently over
vertical distances of a scale height or more.  To understand
the vertical scale of the cloud layer, consider a single condensable
species that is saturated.  From the Clausius-Clapeyron equation, it
can be shown that the ratio of the scale height of the condensable vapor, $H_v$,
to the atmospheric pressure scale height, $H$, is 
\begin{equation}
{H_v\over H} \approx {R_v T\over L\,\Delta \ln T_{\rm sh}}
\label{vapor-scale-height}
\end{equation}
where $R_v$ and $L$ are the specific gas constant and latent heat
of the condensable species, $T$ is temperature, and $\Delta\ln T_{\rm sh}$
is the temperature change of the background temperature profile over a scale height.
For a dry adiabat, $\Delta\ln T_{\rm sh} = R/c_p$, where $R$ and $c_p$ are
the specific gas constant and specific heat of the background air. Adopting
parameters for iron condensation ($R_v = 149\rm\,J\,kg^{-1}\,K^{-1}$ and
$L\approx 6\times10^6\rm\,J\,kg^{-1}$), $T\approx1000\rm\,K$, along with
$\Delta \ln T_{\rm sh} = R/c_p \approx 2/7$ for a diatomic, H$_2$ atmosphere,
we obtain $H_v/H \approx 0.09$.  Therefore, if the temperature profile
is close to an adiabat, the condensation layer is extremely thin compared
to a scale height.  Particles that condense in this layer will gravitationally
settle; particles that settle faster than the upward vertical transport
rate will be confined to a vertically thin layer, but particles that 
settle more slowly than the upward vertical transport rate may be advected
into a vertically extended haze layer.  Nevertheless, if the background
temperature profile is closer to an isotherm, Equation~(\ref{vapor-scale-height})
indicates that even the condensation layer itself could become vertically
extended.

These arguments suggest the following picture for the dust cycle in
the presence of a large-scale atmospheric circulation.  In regions of
ascent, coherent upward motions will bring fresh material upward to
the condensation level, leading to condensation and dust formation.
Extremely small particles could be advected upward with the flow, but
large particles will gravitationally settle, leading to a vertically
thin cloud deck in the ascending regions.  The gas in the ascending
regions will therefore become depleted in condensate as it rises
beyond the condensation level.  In the upper part of the circulation,
this ascending air will presumably move horizontally and begin to
descend (Figure~\ref{wave-driven-circ}). Assuming the circulation extends
sufficiently far above the condensation level, the regions of
descending air will therefore be depleted in condensate, leading to
large-scale ``holes'' in the cloud deck---at least in the horizontal
distribution of large particles.  In contrast, small particles that
have settling times longer than the circulation times may form a
quasi-ubiquitous, unbroken haze layer.  As a result, the particle size
distribution may vary significantly not only vertically but
horizontally as well.  Particles advected downward with the flow in
the descending regions---or that are sufficiently large to settle in
the ascending regions, despite the upward airflow there---will
sublimate, leading to an overall mass balance between condensation and
sublimation.  Even though we are emphasizing here the large-scale
organization, the dust layer could also exhibit patchiness at small
scales, particularly if latent heating is dynamically important (as
occurs in cumulus clouds on Earth).

Jupiter and Saturn represent prime examples of exactly this type of
cycle, with ammonia serving as the condensate.  Jupiter's bright cloud
bands, called ``zones,'' are regions of large-scale ascent at and
above the ammonia cloud deck, and exhibit large mean cloud opacity.
Jupiter's darker cloud bands, called ``belts,'' are regions of
large-scale descent at and above the clouds \citep[e.g.][]
{gierasch-etal-1986, carlson-etal-1994}, and exhibit smaller mean
cloud opacity. Anticyclonic vortices like the Great Red Spot also
appear to be regions of mean large-scale ascent and exhibit a cloud
deck with greater optical depth, and extending to greater altitude,
than the cloud deck in surrounding regions.  At horizontal scales of
$10^3$--$10^4\rm\,km$, Jupiter and Saturn's cloud decks are quite
patchy in particles several microns or larger---as evident in 5-$\mu$m
images of either planet---but relatively more uniform in particles
smaller than a micron \citep[e.g.,][]{west-etal-1986}.  Both belts and
zones exhibit significant small-scale patchiness superposed on top of
this large-scale structure \citep{banfield-etal-1998}.

The large-amplitude lightcurve variations observed by
\citet{artigau-etal-2009} and \citet{radigan-etal-2012} exhibit not
only variation on rotational timescales but also qualitative changes
to the {\it shape} of the rotational modulation over many rotation
periods.  This motivates a consideration of the range of dynamical
timescales operating in the atmosphere of brown dwarfs.  Given our
predicted vertical velocities, and typical horizontal scales of $L\sim
10^4\rm\,$km, the horizontal advection time, $\tau_{\rm adv} \sim L/U
\sim 10^5$--$10^6\rm\,s$.  For a typical brown-dwarf rotation period
of $\sim$$10^4\rm\,s$, this suggests that qualitative, order-unity
changes to the circulation (e.g., in the locations or shapes of
vortices and detailed structure of the turbulence) occur on timescales
of typically $\sim$10--100 rotation periods.  This is qualitatively
consistent with the timescales over which major changes to the
lightcurves are seen in observations
\citep[e.g.,][]{artigau-etal-2009}.

\subsection{Vertical mixing}

A variety of brown dwarfs, including Gl 229B, Gl 570D, HD 3561B, and
2MASS J0415-0935, as well as directly imaged planets such as those
orbiting HR 8799 and 2M1207, show evidence for disequilibrium
abundances of CO, CH$_4$, and/or NH$_3$
\citep[e.g.,][]{noll-etal-1997, saumon-etal-2000, saumon-etal-2006,
  saumon-etal-2007, stephens-etal-2009, leggett-etal-2007,
  leggett-etal-2007b, leggett-etal-2008, hinz-etal-2010,
  barman-etal-2011, barman-etal-2011b, marley-etal-2012}.
Interestingly, in some of these cases, models suggest that the
quenching of CO occurs in the radiative zone rather than the
convection zone, implying that vertical mixing in the stratified
atmosphere is necessary to explain the observations.  Generally, these
1D models parameterize the mixing as an eddy diffusion and tune the
diffusion coefficient to provide a match between synthetic and
observational IR spectra.  Recently, \citet{visscher-moses-2011}
showed using updated reaction kinetics and estimates of quench
timescales that several of these studies underestimated the eddy
diffusivities necessary to explain the inferred CO abundances; 
as a case study, they showed that an eddy diffusivity exceeding
$\sim$$10^2\rm\,m^2\,s^{-1}$ is necessary to explain the CO abundance
on Gl 229B.

\citet{freytag-etal-2010} demonstrate one mechanism for mixing
in the stratified atmosphere, which is the
vertical mixing induced by breaking gravity waves.  Our models in
Section~\ref{atmosphere} suggest an alternate mechanism: that
the mixing results from the large-scale, wave-driven circulation.
Our models predict transport times over a scale height of 
$\tau_{\rm vert}\sim$$10^5$--$10^6\rm\,s$ for a typical L/T dwarf
(see Figure~\ref{windtheory}).  As described in section~\ref{atmosphere},
these motions are not diffusive but comprise vertically coherent
advection extending potentially across several scale heights.
At the temperatures appropriate for typical L/T-dwarf atmospheres,
CO$\rightleftharpoons$CH$_4$ chemical interconversion timescales at low
pressure exceed these transport timescales \citep{cooper-showman-2006,
visscher-moses-2011}, indicating that the
wave-driven circulation can induce chemical quenching of CO and CH$_4$.

The key point is that waves can cause two distinct mechanisms
of vertical transport: direct vertical mixing associated with the wave breaking,
and vertical transport by a large-scale circulation driven by wave interactions
with the mean flow.  Which is larger depends on the wave spectrum, the
detailed structure of the horizontal wind, the atmospheric stratification,
and other properties.  In Earth's stratosphere and lower mesosphere,
vertical mixing by the large-scale (meridional) overturning circulation dominates
over that caused by wave breaking \citep{holton-1983, holton-schoeberl-1988}, but
the reverse may be true at even higher altitudes, in the upper mesosphere.
Detailed calculations will be necessary to evaluate the relative
transport efficiency of the two mechanisms in the context of brown dwarfs.

\section{Conclusions}
\label{conclusions}

We have demonstrated that brown dwarfs should exhibit vigorous
atmospheric circulations at regional to global scales, and that this
circulation will shape the cloudiness, lightcurve variability, and
vertical mixing inferred in their atmospheres.   Our main findings are
as follows:
\begin{itemize}
\item Given inferred rotation rates and plausible wind speeds, the
  circulation on cool brown dwarfs at scales larger than thousands of
  km will be rotationally dominated, exhibiting a horizontal force
  balance between pressure gradient and Coriolis forces (geostrophic
  balance).  The rapid rotation will control many aspects of the
  dynamical behavior at large scales.

\item Using standard vorticity and angular-momentum arguments, we
  showed that, in the convective interior, the rapid rotation and
  nearly isentropic conditions will lead to velocities and convective
  structure aligned---at large scales---along columns parallel to the
  rotation axis.  We presented global, three-dimensional, anelastic
  numerical simulations of convection in brown-dwarf interiors that
  confirm this prediction.  At large scales, convective velocities in
  the deep interior are affected significantly by the rotation,
  leading to a different prediction for the scaling of convective
  velocities with heat flux than predicted by standard (non-rotating)
  mixing length theory.

\item Our 3D convection models show that convection occurs more
  efficiently at high latitudes than at low latitudes, and this can
  lead to a systematic equator-to-pole temperature difference in the
  upper convective zone that may reach $\sim$1 K.  The convection
  develops significant regional-scale structure likely to affect the
  overlying atmosphere in an observable manner.  Scaling arguments
  suggest that, due to Lorentz-force braking in the deep, electrically
  conducting regions, and due to modest thermal-wind shear in the
  outermost (molecular) part of the convection zone, the horizontal
  winds in the convection zone at large scales will be modest
  (probably less than $\sim$$10^2\rm\,m\,s^{-1}$).

\item 
The interaction of convection with the overlying, stably stratified
atmosphere will generate a wealth of atmospheric waves, and we argue
that, just as in the stratospheres of planets in the solar system, the
interaction of these waves with the mean flow will lead to a
significant atmospheric circulation at regional to global scales.  At
horizontal scales exceeding thousands of km, such a wave-driven
circulation will, to good approximation, be geostrophically balanced
and may consist of large-scale stratified turbulence, vortices, and/or
jets.  The geostrophic flow will be accompanied by a large-scale
overturning (``meridional'') circulation that acts to maintain
geostrophy.

\item We presented a simple analytic theory of this wave-driven
  atmospheric circulation which illuminates the dynamical mechanisms
  and enables order-of-magnitude estimates for the characteristic
  horizontal temperature differences, wind speeds, and vertical mixing
  rates as a function of the amplitude of the wave driving, the mean
  stratification of the atmosphere, and other properties.  For
  plausible wave-driving efficiencies, this theory predicts that, at
  photospheric pressures, the circulation will comprise horizontal
  temperature variations of several to $\sim$$50\rm\,K$, horizontal
  wind speeds of $\sim$10--$300\rm\,m\,s^{-1}$, and vertical
  velocities that advect air over a scale height in timescales of
  $\sim$$10^5$--$10^6\rm\,s$.

\item These models help to explain recent evidence for atmospheric
  circulation on cool brown dwarfs. Our predicted large-scale vertical
  motion in the stratified atmosphere would lead to cloud formation and
  mix chemical constituents vertically, helping to explain the
  quenching of CO, CH$_4$, and NH$_3$ inferred on many brown dwarfs.
  This overturning circulation is {\it spatially organized,}
  consisting of coherent, large-scale regions of ascent and descent,
  indicating the likelihood of cloud patchiness that could
  explain lightcurve variability \citep{artigau-etal-2009,
    radigan-etal-2012, buenzli-etal-2012}.  Wind speeds of plausibly
  tens to hundreds of $\rm m\,s^{-1}$ would advect cloudy and
  cloud-free regions horizontally, leading to significant changes in
  the {\it shape} of lightcurves over typically tens of rotation
  periods or more.
\end{itemize}

Our models are of course highly idealized and there remains a pressing
need for detailed, three-dimensional simulations of the atmospheric circulation
and its interaction with the convective interior.  Coupling of such
models to prescriptions of radiative transfer and cloud formation
will be necessary not only to explain the particular shapes and amplitudes
 of brown-dwarf lightcurves as a function of wavelength
but also to understand the role of the atmospheric circulation in influencing
the L/T transition.   Future observations and models are likely
to lead to major new insights over the next few years.


\acknowledgements
This research was supported by Planetary Atmospheres grant
NNX10AB91G and Origins grant NNX12AI79G to APS and by
a Marie Curie Career Integration grant (CIG-304202) to YK.  
APS thanks the Helen Kimel Center for Planetary Science at the
Weizmann Institute of Science for financial support and
a stimulating environment in which to initiate this project.



\bibliographystyle{apj}
\bibliography{showman-bib}

\begin{thebibliography}{128}
\expandafter\ifx\csname natexlab\endcsname\relax\def\natexlab#1{#1}\fi

\bibitem[{{Ackerman} \& {Marley}(2001)}]{ackerman-marley-2001}
{Ackerman}, A.~S., \& {Marley}, M.~S. 2001, \apj, 556, 872

\bibitem[{{Adcroft} {et~al.}(2004){Adcroft}, {Campin}, {Hill}, \&
  {Marshall}}]{adcroft-etal-2004}
{Adcroft}, A., {Campin}, J.-M., {Hill}, C., \& {Marshall}, J. 2004, Monthly
  Weather Review, 132, 2845

\bibitem[{{Allard} {et~al.}(2001){Allard}, {Hauschildt}, {Alexander},
  {Tamanai}, \& {Schweitzer}}]{allard-etal-2001}
{Allard}, F., {Hauschildt}, P.~H., {Alexander}, D.~R., {Tamanai}, A., \&
  {Schweitzer}, A. 2001, \apj, 556, 357

\bibitem[{{Andrews} {et~al.}(1987){Andrews}, {Holton}, \&
  {Leovy}}]{andrews-etal-1987}
{Andrews}, D.~G., {Holton}, J.~R., \& {Leovy}, C.~B. 1987, Middle Atmosphere
  Dynamics (Academic Press, New York)

\bibitem[{{Arakawa} \& {Lamb}(1977)}]{arakawa-lamb-1977}
{Arakawa}, A., \& {Lamb}, V. 1977, Methods in Computational Physics, 17, 173

\bibitem[{{Artigau} {et~al.}(2009){Artigau}, {Bouchard}, {Doyon}, \&
  {Lafreni{\`e}re}}]{artigau-etal-2009}
{Artigau}, {\'E}., {Bouchard}, S., {Doyon}, R., \& {Lafreni{\`e}re}, D. 2009,
  \apj, 701, 1534

\bibitem[{{Aurnou} \& {Olson}(2001)}]{aurnou-olson-2001}
{Aurnou}, J.~M., \& {Olson}, P.~L. 2001, Geophys. Res. Lett., 28, 2557

\bibitem[{{Baldwin} {et~al.}(2001){Baldwin}, {Gray}, {Dunkerton}, {Hamilton},
  {Haynes}, {Randel}, {Holton}, {Alexander}, {Hirota}, {Horinouchi}, {Jones},
  {Kinnersley}, {Marquardt}, {Sato}, \& {Takahashi}}]{baldwin-etal-2001}
{Baldwin}, M.~P., {Gray}, L.~J., {Dunkerton}, T.~J., {Hamilton}, K., {Haynes},
  P.~H., {Randel}, W.~J., {Holton}, J.~R., {Alexander}, M.~J., {Hirota}, I.,
  {Horinouchi}, T., {Jones}, D.~B.~A., {Kinnersley}, J.~S., {Marquardt}, C.,
  {Sato}, K., \& {Takahashi}, M. 2001, Reviews of Geophysics, 39, 179

\bibitem[{{Banfield} {et~al.}(1998){Banfield}, {Gierasch}, {Bell}, {Ustinov},
  {Ingersoll}, {Vasavada}, {West}, \& {Belton}}]{banfield-etal-1998}
{Banfield}, D., {Gierasch}, P.~J., {Bell}, M., {Ustinov}, E., {Ingersoll},
  A.~P., {Vasavada}, A.~R., {West}, R.~A., \& {Belton}, M.~J.~S. 1998, Icarus,
  135, 230

\bibitem[{{Barman} {et~al.}(2011{\natexlab{a}}){Barman}, {Macintosh},
  {Konopacky}, \& {Marois}}]{barman-etal-2011}
{Barman}, T.~S., {Macintosh}, B., {Konopacky}, Q.~M., \& {Marois}, C.
  2011{\natexlab{a}}, \apj, 733, 65

\bibitem[{{Barman} {et~al.}(2011{\natexlab{b}}){Barman}, {Macintosh},
  {Konopacky}, \& {Marois}}]{barman-etal-2011b}
---. 2011{\natexlab{b}}, \apjl, 735, L39

\bibitem[{{B{\'e}zard} {et~al.}(2002){B{\'e}zard}, {Lellouch}, {Strobel},
  {Maillard}, \& {Drossart}}]{bezard-etal-2002}
{B{\'e}zard}, B., {Lellouch}, E., {Strobel}, D., {Maillard}, J.-P., \&
  {Drossart}, P. 2002, Icarus, 159, 95

\bibitem[{{Boubnov} \& {Golitsyn}(1990)}]{boubnov-golitsyn-1990}
{Boubnov}, B.~M., \& {Golitsyn}, G.~S. 1990, Journal of Fluid Mechanics, 219,
  215

\bibitem[{{Bowler} {et~al.}(2010){Bowler}, {Liu}, {Dupuy}, \&
  {Cushing}}]{bowler-etal-2010}
{Bowler}, B.~P., {Liu}, M.~C., {Dupuy}, T.~J., \& {Cushing}, M.~C. 2010, \apj,
  723, 850

\bibitem[{{Buenzli} {et~al.}(2012){Buenzli}, {Apai}, {Morley}, {Flateau},
  {Showman}, {Burrows}, {Marley}, {Lewis}, \& {Reid}}]{buenzli-etal-2012}
{Buenzli}, E., {Apai}, D., {Morley}, C.~V., {Flateau}, D., {Showman}, A.~P.,
  {Burrows}, A., {Marley}, M.~S., {Lewis}, N.~K., \& {Reid}, I.~N. 2012, ArXiv
  e-prints

\bibitem[{{Burgasser} {et~al.}(2002){Burgasser}, {Marley}, {Ackerman},
  {Saumon}, {Lodders}, {Dahn}, {Harris}, \&
  {Kirkpatrick}}]{burgasser-etal-2002}
{Burgasser}, A.~J., {Marley}, M.~S., {Ackerman}, A.~S., {Saumon}, D.,
  {Lodders}, K., {Dahn}, C.~C., {Harris}, H.~C., \& {Kirkpatrick}, J.~D. 2002,
  \apjl, 571, L151

\bibitem[{{Burrows} {et~al.}(2001){Burrows}, {Hubbard}, {Lunine}, \&
  {Liebert}}]{burrows-etal-2001}
{Burrows}, A., {Hubbard}, W.~B., {Lunine}, J.~I., \& {Liebert}, J. 2001,
  Reviews of Modern Physics, 73, 719

\bibitem[{{Burrows} {et~al.}(1997){Burrows}, {Marley}, {Hubbard}, {Lunine},
  {Guillot}, {Saumon}, {Freedman}, {Sudarsky}, \& {Sharp}}]{burrows-etal-1997}
{Burrows}, A., {Marley}, M., {Hubbard}, W.~B., {Lunine}, J.~I., {Guillot}, T.,
  {Saumon}, D., {Freedman}, R., {Sudarsky}, D., \& {Sharp}, C. 1997, \apj, 491,
  856

\bibitem[{{Burrows} {et~al.}(2006{\natexlab{a}}){Burrows}, {Sudarsky}, \&
  {Hubeny}}]{burrows-etal-2006b}
{Burrows}, A., {Sudarsky}, D., \& {Hubeny}, I. 2006{\natexlab{a}}, \apj, 640,
  1063

\bibitem[{{Burrows} {et~al.}(2006{\natexlab{b}}){Burrows}, {Sudarsky}, \&
  {Hubeny}}]{burrows-etal-2006}
---. 2006{\natexlab{b}}, \apj, 650, 1140

\bibitem[{{Busse}(1976)}]{busse-1976}
{Busse}, F.~H. 1976, Icarus, 29, 255

\bibitem[{{Busse}(2002)}]{busse-2002}
---. 2002, Physics of Fluids, 14, 1301

\bibitem[{{Carlson} {et~al.}(1994){Carlson}, {Lacis}, \&
  {Rossow}}]{carlson-etal-1994}
{Carlson}, B.~E., {Lacis}, A.~A., \& {Rossow}, W.~B. 1994, \jgr, 99, 14623

\bibitem[{{Chabrier} {et~al.}(2000){Chabrier}, {Baraffe}, {Allard}, \&
  {Hauschildt}}]{chabrier-etal-2000}
{Chabrier}, G., {Baraffe}, I., {Allard}, F., \& {Hauschildt}, P. 2000, \apj,
  542, 464

\bibitem[{{Cho} \& {Polvani}(1996)}]{cho-polvani-1996a}
{Cho}, J.~Y.-K., \& {Polvani}, L.~M. 1996, Science, 8, 1

\bibitem[{{Christensen}(2001)}]{christensen-2001}
{Christensen}, U.~R. 2001, Geophys. Res. Lett., 28, 2553

\bibitem[{{Christensen}(2002)}]{christensen-2002}
---. 2002, Journal of Fluid Mechanics, 470, 115

\bibitem[{{Christensen}(2010)}]{christensen-2010}
---. 2010, \ssr, 152, 565

\bibitem[{{Christensen} \& {Aubert}(2006)}]{christensen-aubert-2006}
{Christensen}, U.~R., \& {Aubert}, J. 2006, Geophysical Journal International,
  166, 97

\bibitem[{{Christensen} {et~al.}(2009){Christensen}, {Holzwarth}, \&
  {Reiners}}]{christensen-etal-2009}
{Christensen}, U.~R., {Holzwarth}, V., \& {Reiners}, A. 2009, \nat, 457, 167

\bibitem[{{Clayton}(1968)}]{clayton-1968}
{Clayton}, D.~D. 1968, Principles of Stellar Evolution and Nucleosynthesis
  (McGraw-Hill, New York)

\bibitem[{{Cooper} \& {Showman}(2006)}]{cooper-showman-2006}
{Cooper}, C.~S., \& {Showman}, A.~P. 2006, \apj, 649, 1048

\bibitem[{{Currie} {et~al.}(2011){Currie}, {Burrows}, {Itoh}, {Matsumura},
  {Fukagawa}, {Apai}, {Madhusudhan}, {Hinz}, {Rodigas}, {Kasper}, {Pyo}, \&
  {Ogino}}]{currie-etal-2011}
{Currie}, T., {Burrows}, A., {Itoh}, Y., {Matsumura}, S., {Fukagawa}, M.,
  {Apai}, D., {Madhusudhan}, N., {Hinz}, P.~M., {Rodigas}, T.~J., {Kasper}, M.,
  {Pyo}, T.-S., \& {Ogino}, S. 2011, \apj, 729, 128

\bibitem[{{Cushing} {et~al.}(2011){Cushing}, {Kirkpatrick}, {Gelino},
  {Griffith}, {Skrutskie}, {Mainzer}, {Marsh}, {Beichman}, {Burgasser},
  {Prato}, {Simcoe}, {Marley}, {Saumon}, {Freedman}, {Eisenhardt}, \&
  {Wright}}]{cushing-etal-2011}
{Cushing}, M.~C., {Kirkpatrick}, J.~D., {Gelino}, C.~R., {Griffith}, R.~L.,
  {Skrutskie}, M.~F., {Mainzer}, A., {Marsh}, K.~A., {Beichman}, C.~A.,
  {Burgasser}, A.~J., {Prato}, L.~A., {Simcoe}, R.~A., {Marley}, M.~S.,
  {Saumon}, D., {Freedman}, R.~S., {Eisenhardt}, P.~R., \& {Wright}, E.~L.
  2011, \apj, 743, 50

\bibitem[{{Cushing} {et~al.}(2006){Cushing}, {Roellig}, {Marley}, {Saumon},
  {Leggett}, {Kirkpatrick}, {Wilson}, {Sloan}, {Mainzer}, {Van Cleve}, \&
  {Houck}}]{cushing-etal-2006}
{Cushing}, M.~C., {Roellig}, T.~L., {Marley}, M.~S., {Saumon}, D., {Leggett},
  S.~K., {Kirkpatrick}, J.~D., {Wilson}, J.~C., {Sloan}, G.~C., {Mainzer},
  A.~K., {Van Cleve}, J.~E., \& {Houck}, J.~R. 2006, \apj, 648, 614

\bibitem[{{Dritschel} \& {McIntyre}(2008)}]{dritschel-mcintyre-2008}
{Dritschel}, D.~G., \& {McIntyre}, M.~E. 2008, Journal of Atmospheric Sciences,
  65, 855

\bibitem[{{Dritschel} \& {Scott}(2011)}]{dritschel-scott-2011}
{Dritschel}, D.~G., \& {Scott}, R.~K. 2011, Phil. Trans. Roy. Soc. A, 369, 754

\bibitem[{{Fegley} \& {Lodders}(1996)}]{fegley-lodders-1996}
{Fegley}, Jr., B., \& {Lodders}, K. 1996, \apjl, 472, L37

\bibitem[{{Fernando} {et~al.}(1991){Fernando}, {Chen}, \&
  {Boyer}}]{fernando-etal-1991}
{Fernando}, H.~J.~S., {Chen}, R.-R., \& {Boyer}, D.~L. 1991, Journal of Fluid
  Mechanics, 228, 513

\bibitem[{{Freytag} {et~al.}(2010){Freytag}, {Allard}, {Ludwig}, {Homeier}, \&
  {Steffen}}]{freytag-etal-2010}
{Freytag}, B., {Allard}, F., {Ludwig}, H.-G., {Homeier}, D., \& {Steffen}, M.
  2010, \aap, 513, A19

\bibitem[{{Friedson}(1999)}]{friedson-1999}
{Friedson}, A.~J. 1999, Icarus, 137, 34

\bibitem[{{Galicher} {et~al.}(2011){Galicher}, {Marois}, {Macintosh}, {Barman},
  \& {Konopacky}}]{galicher-etal-2011}
{Galicher}, R., {Marois}, C., {Macintosh}, B., {Barman}, T., \& {Konopacky}, Q.
  2011, \apjl, 739, L41

\bibitem[{{Gierasch} {et~al.}(1986){Gierasch}, {Magalhaes}, \&
  {Conrath}}]{gierasch-etal-1986}
{Gierasch}, P.~J., {Magalhaes}, J.~A., \& {Conrath}, B.~J. 1986, Icarus, 67,
  456

\bibitem[{{Gilman} \& {Glatzmaier}(1981)}]{gilman-glatzmaier-1981}
{Gilman}, P.~A., \& {Glatzmaier}, G.~A. 1981, \apjs, 45, 335

\bibitem[{{Glatzmaier} {et~al.}(2009){Glatzmaier}, {Evonuk}, \&
  {Rogers}}]{glatzmaier-etal-2009}
{Glatzmaier}, G.~A., {Evonuk}, M., \& {Rogers}, T.~M. 2009, Geophys. Astrophy.
  Fluid Dyn., 103, 31

\bibitem[{{Goldreich} \& {Kumar}(1990)}]{goldreich-kumar-1990}
{Goldreich}, P., \& {Kumar}, P. 1990, \apj, 363, 694

\bibitem[{{Golitsyn}(1980)}]{golitsyn-1980}
{Golitsyn}, G.~S. 1980, Doklady Akademii Nauk SSSR, 251, 1356

\bibitem[{{Golitsyn}(1981)}]{golitsyn-1981}
---. 1981, Doklady Akademii Nauk SSSR, 261, 317

\bibitem[{{Griffith} \& {Yelle}(1999)}]{griffith-yelle-1999}
{Griffith}, C.~A., \& {Yelle}, R.~V. 1999, \apjl, 519, L85

\bibitem[{{Grote} {et~al.}(2000){Grote}, {Busse}, \&
  {Tilgner}}]{grote-etal-2000a}
{Grote}, E., {Busse}, F.~H., \& {Tilgner}, A. 2000, Physics of the Earth and
  Planetary Interiors, 117, 259

\bibitem[{{Guillot} \& {Morel}(1995)}]{guillot-morel-1995}
{Guillot}, T., \& {Morel}, P. 1995, \aaps, 109, 109

\bibitem[{{Guillot} {et~al.}(2004){Guillot}, {Stevenson}, {Hubbard}, \&
  {Saumon}}]{guillot-etal-2004}
{Guillot}, T., {Stevenson}, D.~J., {Hubbard}, W.~B., \& {Saumon}, D. 2004, in
  Jupiter: The Planet, Satellites and Magnetosphere, ed. F.~{Bagenal}, T.~E.
  {Dowling}, \& W.~B. {McKinnon} (Cambridge Univ. Press), 35--57

\bibitem[{{Haynes}(2005)}]{haynes-2005}
{Haynes}, P. 2005, Annual Review of Fluid Mechanics, 37, 263

\bibitem[{{Haynes} {et~al.}(1991){Haynes}, {McIntyre}, {Shepherd}, {Marks}, \&
  {Shine}}]{haynes-etal-1991}
{Haynes}, P.~H., {McIntyre}, M.~E., {Shepherd}, T.~G., {Marks}, C.~J., \&
  {Shine}, K.~P. 1991, Journal of Atmospheric Sciences, 48, 651

\bibitem[{{Heimpel} {et~al.}(2005){Heimpel}, {Aurnou}, \&
  {Wicht}}]{heimpel-etal-2005}
{Heimpel}, M., {Aurnou}, J., \& {Wicht}, J. 2005, \nat, 438, 193

\bibitem[{{Hinz} {et~al.}(2010){Hinz}, {Rodigas}, {Kenworthy}, {Sivanandam},
  {Heinze}, {Mamajek}, \& {Meyer}}]{hinz-etal-2010}
{Hinz}, P.~M., {Rodigas}, T.~J., {Kenworthy}, M.~A., {Sivanandam}, S.,
  {Heinze}, A.~N., {Mamajek}, E.~E., \& {Meyer}, M.~R. 2010, \apj, 716, 417

\bibitem[{{Holton}(1983)}]{holton-1983}
{Holton}, J.~R. 1983, Journal of Atmospheric Sciences, 40, 2497

\bibitem[{{Holton}(2004)}]{holton-2004}
---. 2004, An Introduction to Dynamic Meteorology, 4th Ed. (Academic Press, San
  Diego)

\bibitem[{{Holton} \& {Schoeberl}(1988)}]{holton-schoeberl-1988}
{Holton}, J.~R., \& {Schoeberl}, M.~R. 1988, \jgr, 93, 11075

\bibitem[{{Huang} \& {Robinson}(1998)}]{huang-robinson-1998}
{Huang}, H.-P., \& {Robinson}, W.~A. 1998, Journal of Atmospheric Sciences, 55,
  611

\bibitem[{{Hubeny} \& {Burrows}(2007)}]{hubeny-burrows-2007}
{Hubeny}, I., \& {Burrows}, A. 2007, \apj, 669, 1248

\bibitem[{{Ingersoll}(1990)}]{ingersoll-1990}
{Ingersoll}, A.~P. 1990, Science, 248, 308

\bibitem[{{Ingersoll} \& {Pollard}(1982)}]{ingersoll-pollard-1982}
{Ingersoll}, A.~P., \& {Pollard}, D. 1982, Icarus, 52, 62

\bibitem[{{James}(1994)}]{james-1994}
{James}, I.~N. 1994, Introduction to Circulating Atmospheres (Cambridge
  Atmospheric and Space Science Series, Cambridge University Press, UK)

\bibitem[{{Jones} \& {Kuzanyan}(2009)}]{jones-kuzanyan-2009}
{Jones}, C.~A., \& {Kuzanyan}, K.~M. 2009, Icarus, 204, 227

\bibitem[{{Kaspi}(2008)}]{kaspi-2008}
{Kaspi}, Y. 2008, PhD thesis, Massachusetts Institute of Technology

\bibitem[{{Kaspi} {et~al.}(2009){Kaspi}, {Flierl}, \&
  {Showman}}]{kaspi-etal-2009}
{Kaspi}, Y., {Flierl}, G.~R., \& {Showman}, A.~P. 2009, Icarus, 202, 525

\bibitem[{{Kirk} \& {Stevenson}(1987)}]{kirk-stevenson-1987}
{Kirk}, R.~L., \& {Stevenson}, D.~J. 1987, \apj, 316, 836

\bibitem[{{Kirkpatrick}(2005)}]{kirkpatrick-2005}
{Kirkpatrick}, J.~D. 2005, \araa, 43, 195

\bibitem[{{Kirkpatrick} {et~al.}(1999){Kirkpatrick}, {Reid}, {Liebert},
  {Cutri}, {Nelson}, {Beichman}, {Dahn}, {Monet}, {Gizis}, \&
  {Skrutskie}}]{kirkpatrick-etal-1999}
{Kirkpatrick}, J.~D., {Reid}, I.~N., {Liebert}, J., {Cutri}, R.~M., {Nelson},
  B., {Beichman}, C.~A., {Dahn}, C.~C., {Monet}, D.~G., {Gizis}, J.~E., \&
  {Skrutskie}, M.~F. 1999, \apj, 519, 802

\bibitem[{{Knapp} {et~al.}(2004){Knapp}, {Leggett}, {Fan}, {Marley}, {Geballe},
  {Golimowski}, {Finkbeiner}, {Gunn}, {Hennawi}, {Ivezi{\'c}}, {Lupton},
  {Schlegel}, {Strauss}, {Tsvetanov}, {Chiu}, {Hoversten}, {Glazebrook},
  {Zheng}, {Hendrickson}, {Williams}, {Uomoto}, {Vrba}, {Henden}, {Luginbuhl},
  {Guetter}, {Munn}, {Canzian}, {Schneider}, \& {Brinkmann}}]{knapp-etal-2004}
{Knapp}, G.~R., {Leggett}, S.~K., {Fan}, X., {Marley}, M.~S., {Geballe}, T.~R.,
  {Golimowski}, D.~A., {Finkbeiner}, D., {Gunn}, J.~E., {Hennawi}, J.,
  {Ivezi{\'c}}, Z., {Lupton}, R.~H., {Schlegel}, D.~J., {Strauss}, M.~A.,
  {Tsvetanov}, Z.~I., {Chiu}, K., {Hoversten}, E.~A., {Glazebrook}, K.,
  {Zheng}, W., {Hendrickson}, M., {Williams}, C.~C., {Uomoto}, A., {Vrba},
  F.~J., {Henden}, A.~A., {Luginbuhl}, C.~B., {Guetter}, H.~H., {Munn}, J.~A.,
  {Canzian}, B., {Schneider}, D.~P., \& {Brinkmann}, J. 2004, \aj, 127, 3553

\bibitem[{{Leggett} {et~al.}(2010){Leggett}, {Burningham}, {Saumon}, {Marley},
  {Warren}, {Smart}, {Jones}, {Lucas}, {Pinfield}, \&
  {Tamura}}]{leggett-etal-2010}
{Leggett}, S.~K., {Burningham}, B., {Saumon}, D., {Marley}, M.~S., {Warren},
  S.~J., {Smart}, R.~L., {Jones}, H.~R.~A., {Lucas}, P.~W., {Pinfield}, D.~J.,
  \& {Tamura}, M. 2010, \apj, 710, 1627

\bibitem[{{Leggett} {et~al.}(2007{\natexlab{a}}){Leggett}, {Marley},
  {Freedman}, {Saumon}, {Liu}, {Geballe}, {Golimowski}, \&
  {Stephens}}]{leggett-etal-2007b}
{Leggett}, S.~K., {Marley}, M.~S., {Freedman}, R., {Saumon}, D., {Liu}, M.~C.,
  {Geballe}, T.~R., {Golimowski}, D.~A., \& {Stephens}, D.~C.
  2007{\natexlab{a}}, \apj, 667, 537

\bibitem[{{Leggett} {et~al.}(2008){Leggett}, {Saumon}, {Albert}, {Cushing},
  {Liu}, {Luhman}, {Marley}, {Kirkpatrick}, {Roellig}, \&
  {Allers}}]{leggett-etal-2008}
{Leggett}, S.~K., {Saumon}, D., {Albert}, L., {Cushing}, M.~C., {Liu}, M.~C.,
  {Luhman}, K.~L., {Marley}, M.~S., {Kirkpatrick}, J.~D., {Roellig}, T.~L., \&
  {Allers}, K.~N. 2008, \apj, 682, 1256

\bibitem[{{Leggett} {et~al.}(2007{\natexlab{b}}){Leggett}, {Saumon}, {Marley},
  {Geballe}, {Golimowski}, {Stephens}, \& {Fan}}]{leggett-etal-2007}
{Leggett}, S.~K., {Saumon}, D., {Marley}, M.~S., {Geballe}, T.~R.,
  {Golimowski}, D.~A., {Stephens}, D., \& {Fan}, X. 2007{\natexlab{b}}, \apj,
  655, 1079

\bibitem[{{Liu} {et~al.}(2008){Liu}, {Goldreich}, \&
  {Stevenson}}]{liu-etal-2008}
{Liu}, J., {Goldreich}, P.~M., \& {Stevenson}, D.~J. 2008, Icarus, 196, 653

\bibitem[{{Liu} \& {Schneider}(2010)}]{liu-schneider-2010}
{Liu}, J., \& {Schneider}, T. 2010, Journal of Atmospheric Sciences, 67, 3652

\bibitem[{{Madhusudhan} {et~al.}(2011){Madhusudhan}, {Burrows}, \&
  {Currie}}]{madhusudhan-etal-2011b}
{Madhusudhan}, N., {Burrows}, A., \& {Currie}, T. 2011, \apj, 737, 34

\bibitem[{{Marcus} {et~al.}(2000){Marcus}, {Kundu}, \&
  {Lee}}]{marcus-etal-2000}
{Marcus}, P.~S., {Kundu}, T., \& {Lee}, C. 2000, Physics of Plasmas, 7, 1630

\bibitem[{{Marley} {et~al.}(2012){Marley}, {Saumon}, {Cushing}, {Ackerman},
  {Fortney}, \& {Freedman}}]{marley-etal-2012}
{Marley}, M.~S., {Saumon}, D., {Cushing}, M., {Ackerman}, A.~S., {Fortney},
  J.~J., \& {Freedman}, R. 2012, \apj, 754, 135

\bibitem[{{Marley} {et~al.}(2010){Marley}, {Saumon}, \&
  {Goldblatt}}]{marley-etal-2010}
{Marley}, M.~S., {Saumon}, D., \& {Goldblatt}, C. 2010, \apjl, 723, L117

\bibitem[{{Marley} {et~al.}(1996){Marley}, {Saumon}, {Guillot}, {Freedman},
  {Hubbard}, {Burrows}, \& {Lunine}}]{marley-etal-1996}
{Marley}, M.~S., {Saumon}, D., {Guillot}, T., {Freedman}, R.~S., {Hubbard},
  W.~B., {Burrows}, A., \& {Lunine}, J.~I. 1996, Science, 272, 1919

\bibitem[{{Marley} {et~al.}(2002){Marley}, {Seager}, {Saumon}, {Lodders},
  {Ackerman}, {Freedman}, \& {Fan}}]{marley-etal-2002}
{Marley}, M.~S., {Seager}, S., {Saumon}, D., {Lodders}, K., {Ackerman}, A.~S.,
  {Freedman}, R.~S., \& {Fan}, X. 2002, \apj, 568, 335

\bibitem[{{Miesch} \& {Toomre}(2009)}]{miesch-toomre-2009}
{Miesch}, M.~S., \& {Toomre}, J. 2009, Annual Review of Fluid Mechanics, 41,
  317

\bibitem[{{Moreno} \& {Sedano}(1997)}]{moreno-sedano-1997}
{Moreno}, F., \& {Sedano}, J. 1997, Icarus, 130, 36

\bibitem[{{Nellis}(2000)}]{nellis-2000}
{Nellis}, W.~J. 2000, \planss, 48, 671

\bibitem[{{Nellis}(2006)}]{nellis-2006}
---. 2006, Reports on Progress in Physics, 69, 1479

\bibitem[{{Nellis} {et~al.}(1995){Nellis}, {Ross}, \&
  {Holmes}}]{nellis-etal-1995}
{Nellis}, W.~J., {Ross}, M., \& {Holmes}, N.~C. 1995, Science, 269, 1249

\bibitem[{{Nellis} {et~al.}(1996){Nellis}, {Weir}, \&
  {Mitchell}}]{nellis-etal-1996}
{Nellis}, W.~J., {Weir}, S.~T., \& {Mitchell}, A.~C. 1996, Science, 273, 936

\bibitem[{{Noll} {et~al.}(1997){Noll}, {Geballe}, \& {Marley}}]{noll-etal-1997}
{Noll}, K.~S., {Geballe}, T.~R., \& {Marley}, M.~S. 1997, \apjl, 489, L87+

\bibitem[{{Nozawa} \& {Yoden}(1997)}]{nozawa-yoden-1997a}
{Nozawa}, T., \& {Yoden}, S. 1997, Physics of Fluids, 9, 2081

\bibitem[{{Ogura} \& {Phillips}(1962)}]{ogura-phillips-1962}
{Ogura}, Y., \& {Phillips}, N.~A. 1962, Journal of Atmospheric Sciences, 19,
  173

\bibitem[{{Pedlosky}(1987)}]{pedlosky-1987}
{Pedlosky}, J. 1987, Geophysical Fluid Dynamics, 2nd Ed. (Springer-Verlag, New
  York)

\bibitem[{{Peixoto} \& {Oort}(1992)}]{peixoto-oort-1992}
{Peixoto}, J.~P., \& {Oort}, A.~H. 1992, Physics of Climate (American Institute
  of Physics, New York)

\bibitem[{{Prinn} \& {Barshay}(1977)}]{prinn-barshay-1977}
{Prinn}, R.~G., \& {Barshay}, S.~S. 1977, Science, 198, 1031

\bibitem[{{Radigan} {et~al.}(2012){Radigan}, {Jayawardhana}, {Lafreni{\`e}re},
  {Artigau}, {Marley}, \& {Saumon}}]{radigan-etal-2012}
{Radigan}, J., {Jayawardhana}, R., {Lafreni{\`e}re}, D., {Artigau}, {\'E}.,
  {Marley}, M., \& {Saumon}, D. 2012, \apj, 750, 105

\bibitem[{{Reiners} \& {Basri}(2008)}]{reiners-basri-2008}
{Reiners}, A., \& {Basri}, G. 2008, \apj, 684, 1390

\bibitem[{{Reiners} \& {Christensen}(2010)}]{reiners-christensen-2010}
{Reiners}, A., \& {Christensen}, U.~R. 2010, \aap, 522, A13

\bibitem[{{Rogers} {et~al.}(2012){Rogers}, {Lin}, \& {Lau}}]{rogers-etal-2012}
{Rogers}, T.~M., {Lin}, D.~N.~C., \& {Lau}, H.~H.~B. 2012, ApJL, 758, L6

\bibitem[{{Saumon} {et~al.}(1995){Saumon}, {Chabrier}, \& {van
  Horn}}]{saumon-etal-1995}
{Saumon}, D., {Chabrier}, G., \& {van Horn}, H.~M. 1995, \apjs, 99, 713

\bibitem[{{Saumon} {et~al.}(2000){Saumon}, {Geballe}, {Leggett}, {Marley},
  {Freedman}, {Lodders}, {Fegley}, \& {Sengupta}}]{saumon-etal-2000}
{Saumon}, D., {Geballe}, T.~R., {Leggett}, S.~K., {Marley}, M.~S., {Freedman},
  R.~S., {Lodders}, K., {Fegley}, Jr., B., \& {Sengupta}, S.~K. 2000, \apj,
  541, 374

\bibitem[{{Saumon} \& {Marley}(2008)}]{saumon-marley-2008}
{Saumon}, D., \& {Marley}, M.~S. 2008, \apj, 689, 1327

\bibitem[{{Saumon} {et~al.}(2006){Saumon}, {Marley}, {Cushing}, {Leggett},
  {Roellig}, {Lodders}, \& {Freedman}}]{saumon-etal-2006}
{Saumon}, D., {Marley}, M.~S., {Cushing}, M.~C., {Leggett}, S.~K., {Roellig},
  T.~L., {Lodders}, K., \& {Freedman}, R.~S. 2006, \apj, 647, 552

\bibitem[{{Saumon} {et~al.}(2007){Saumon}, {Marley}, {Leggett}, {Geballe},
  {Stephens}, {Golimowski}, {Cushing}, {Fan}, {Rayner}, {Lodders}, \&
  {Freedman}}]{saumon-etal-2007}
{Saumon}, D., {Marley}, M.~S., {Leggett}, S.~K., {Geballe}, T.~R., {Stephens},
  D., {Golimowski}, D.~A., {Cushing}, M.~C., {Fan}, X., {Rayner}, J.~T.,
  {Lodders}, K., \& {Freedman}, R.~S. 2007, \apj, 656, 1136

\bibitem[{{Scott} \& {Polvani}(2007)}]{scott-polvani-2007}
{Scott}, R.~K., \& {Polvani}, L. 2007, J. Atmos. Sci, 64, 3158

\bibitem[{{Shepherd}(2000)}]{shepherd-2000}
{Shepherd}, T.~G. 2000, Journal of Atmospheric and Solar-Terrestrial Physics,
  62, 1587

\bibitem[{{Shepherd}(2003)}]{shepherd-2003}
---. 2003, Chem. Rev., 103, 4509

\bibitem[{{Showman}(2007)}]{showman-2007}
{Showman}, A.~P. 2007, J. Atmos. Sci., 64, 3132

\bibitem[{{Showman} {et~al.}(2010){Showman}, {Cho}, \&
  {Menou}}]{showman-etal-2010}
{Showman}, A.~P., {Cho}, J.~Y.-K., \& {Menou}, K. 2010, {\rm Atmospheric
  circulation of Exoplanets}. {\rm In} {\it Exoplanets} (S. Seager, Ed.) (Univ.
  Arizona Press), 471--516

\bibitem[{{Showman} \& {Guillot}(2002)}]{showman-guillot-2002}
{Showman}, A.~P., \& {Guillot}, T. 2002, \aap, 385, 166

\bibitem[{{Showman} {et~al.}(2011){Showman}, {Kaspi}, \&
  {Flierl}}]{showman-etal-2011}
{Showman}, A.~P., {Kaspi}, Y., \& {Flierl}, G.~R. 2011, Icarus, 211, 1258

\bibitem[{{Simon-Miller} {et~al.}(2006){Simon-Miller}, {Conrath}, {Gierasch},
  {Orton}, {Achterberg}, {Flasar}, \& {Fisher}}]{simon-miller-etal-2006}
{Simon-Miller}, A.~A., {Conrath}, B.~J., {Gierasch}, P.~J., {Orton}, G.~S.,
  {Achterberg}, R.~K., {Flasar}, F.~M., \& {Fisher}, B.~M. 2006, Icarus, 180,
  98

\bibitem[{{Skemer} {et~al.}(2011){Skemer}, {Close}, {Sz{\H u}cs}, {Apai},
  {Pascucci}, \& {Biller}}]{skemer-etal-2011}
{Skemer}, A.~J., {Close}, L.~M., {Sz{\H u}cs}, L., {Apai}, D., {Pascucci}, I.,
  \& {Biller}, B.~A. 2011, \apj, 732, 107

\bibitem[{{Smith}(2004)}]{smith-2004}
{Smith}, K.~S. 2004, J. Atmos. Sciences, 61, 1420

\bibitem[{{Smith} \& {Vallis}(2001)}]{smith-vallis-2001}
{Smith}, K.~S., \& {Vallis}, G.~K. 2001, Journal of Physical Oceanography, 31,
  554

\bibitem[{{Starchenko} \& {Jones}(2002)}]{starchenko-jones-2002}
{Starchenko}, S.~V., \& {Jones}, C.~A. 2002, Icarus, 157, 426

\bibitem[{{Stephens} {et~al.}(2009){Stephens}, {Leggett}, {Cushing}, {Marley},
  {Saumon}, {Geballe}, {Golimowski}, {Fan}, \& {Noll}}]{stephens-etal-2009}
{Stephens}, D.~C., {Leggett}, S.~K., {Cushing}, M.~C., {Marley}, M.~S.,
  {Saumon}, D., {Geballe}, T.~R., {Golimowski}, D.~A., {Fan}, X., \& {Noll},
  K.~S. 2009, \apj, 702, 154

\bibitem[{{Stevenson}(1979)}]{stevenson-1979}
{Stevenson}, D.~J. 1979, Geophysical and Astrophysical Fluid Dynamics, 12, 139

\bibitem[{{Stevenson}(2003)}]{stevenson-2003}
---. 2003, Earth and Planetary Science Letters, 208, 1

\bibitem[{{Stevenson}(2010)}]{stevenson-2010}
---. 2010, \ssr, 152, 651

\bibitem[{{Tsuji}(2002)}]{tsuji-2002}
{Tsuji}, T. 2002, \apj, 575, 264

\bibitem[{{Vallis}(2006)}]{vallis-2006}
{Vallis}, G.~K. 2006, Atmospheric and Oceanic Fluid Dynamics: Fundamentals and
  Large-Scale Circulation (Cambridge Univ. Press, Cambridge, UK)

\bibitem[{{Vasavada} \& {Showman}(2005)}]{vasavada-showman-2005}
{Vasavada}, A.~R., \& {Showman}, A.~P. 2005, Reports of Progress in Physics,
  68, 1935

\bibitem[{{Visscher} \& {Moses}(2011)}]{visscher-moses-2011}
{Visscher}, C., \& {Moses}, J.~I. 2011, \apj, 738, 72

\bibitem[{{Weir} {et~al.}(1996){Weir}, {Mitchell}, \&
  {Nellis}}]{weir-etal-1996}
{Weir}, S.~T., {Mitchell}, A.~C., \& {Nellis}, W.~J. 1996, Physical Review
  Letters, 76, 1860

\bibitem[{{West} {et~al.}(1992){West}, {Friedson}, \&
  {Appleby}}]{west-etal-1992}
{West}, R.~A., {Friedson}, A.~J., \& {Appleby}, J.~F. 1992, Icarus, 100, 245

\bibitem[{{West} {et~al.}(1986){West}, {Strobel}, \&
  {Tomasko}}]{west-etal-1986}
{West}, R.~A., {Strobel}, D.~F., \& {Tomasko}, M.~G. 1986, Icarus, 65, 161

\bibitem[{{Zapatero Osorio} {et~al.}(2006){Zapatero Osorio}, {Mart{\'{\i}}n},
  {Bouy}, {Tata}, {Deshpande}, \& {Wainscoat}}]{zapatero-osorio-etal-2006}
{Zapatero Osorio}, M.~R., {Mart{\'{\i}}n}, E.~L., {Bouy}, H., {Tata}, R.,
  {Deshpande}, R., \& {Wainscoat}, R.~J. 2006, \apj, 647, 1405

\end{thebibliography}



\end{document}